\documentclass[sigconf]{acmart}
\AtBeginDocument{%
  }

\copyrightyear{2026}
\acmYear{2026}
\setcopyright{cc}
\setcctype{by}
\acmConference[DIS '26]{Designing Interactive Systems Conference}{June 13--17, 2026}{Singapore, Singapore}
\acmBooktitle{Designing Interactive Systems Conference (DIS '26), June 13--17, 2026, Singapore, Singapore}
\acmDOI{10.1145/3800645.3813014}
\acmISBN{979-8-4007-2563-0/2026/06}

\usepackage{graphicx}
\usepackage{soul}
\usepackage{xcolor}
\usepackage{multicol}
\usepackage{enumitem}
\definecolor{orangeOrange}{HTML}{E75F24}
\definecolor{pinkPink}{HTML}{E7247C}
\definecolor{blueBlue}{HTML}{008CFF}
\usepackage{hyperref}
\usepackage{booktabs}
\usepackage{array}
\usepackage{tabularx}
\usepackage{placeins}

\begin{document}

\title{PrivacyMotiv: Vulnerability-Centered Persona Journeys for Empathic Privacy Reviews in UX Design}

\author{Zeya Chen}
\authornote{Both authors contributed equally to this work.}
\affiliation{%
  \institution{Northeastern University}
  \city{Boston}
  \state{Massachusetts}
  \country{USA}
}
\affiliation{%
  \institution{Institute of Design (ID) at Illinois Tech}
  \city{Chicago}
  \state{Illinois}
  \country{USA}
}
\email{zey.chen@northeastern.edu}

\author{Jianing Wen}
\authornotemark[1]
\affiliation{%
  \institution{Northeastern University}
  \city{Boston}
  \state{Massachusetts}
  \country{USA}
}
\email{wen.jiani@northeastern.edu}

\author{Yaxing Yao}
\affiliation{%
  \institution{Johns Hopkins University}
  \city{Baltimore}
  \state{Maryland}
  \country{USA}}
\email{yaxing@jhu.edu}

\author{Toby Jia-Jun Li}
\affiliation{%
  \institution{University of Notre Dame}
  \city{Notre Dame}
  \state{Indiana}
  \country{USA}}
\email{toby.j.li@nd.edu}

\author{Tianshi Li}
\affiliation{%
  \institution{Northeastern University}
  \city{Boston}
  \state{Massachusetts}
  \country{USA}}
\email{tia.li@northeastern.edu}

\renewcommand{\shortauthors}{Chen \& Wen et al.}
\newcommand{\systemname}{PrivacyMotiv}

\ccsdesc[500]{Human-centered computing~User studies}
\ccsdesc[500]{Human-centered computing~User interface design}
\ccsdesc[300]{Human-centered computing~Interaction design}
\ccsdesc[500]{Security and privacy~Privacy protections}
\ccsdesc[300]{Security and privacy~Human and societal aspects of security and privacy}
\ccsdesc[300]{Human-centered computing~Empirical studies in HCI}

\begin{abstract}

UX professionals routinely conduct design reviews, yet privacy concerns are often overlooked, not only due to limited tools, but more fundamentally from low intrinsic motivation, driven by limited privacy knowledge, weak empathy for unexpectedly affected users, and low autonomy in identifying harms. We present \textbf{\textit{PrivacyMotiv}}, an LLM-powered system that generates vulnerability-centered personas, persona journey stories, and traceable design diagnoses grounded in lo-fi user flows to support privacy-oriented UX design review. In a within-subjects study with professional UX practitioners (N=16), \textit{PrivacyMotiv} significantly improved empathy, intrinsic motivation, and perceived usefulness, with participants identifying 59\% more privacy issues and proposing 70\% more redesign solutions compared to self-proposed methods. This work contributes empirical insight into motivational barriers in privacy-aware UX and a structured, narrative-driven approach for integrating privacy review into early-stage UX practice.

\end{abstract} 

\keywords{privacy-aware design, UX design review, vulnerability-centered personas, empathy, intrinsic motivation, large language models, user journey, privacy harms, usable privacy}

\begin{teaserfigure}
  \includegraphics[width=\textwidth]{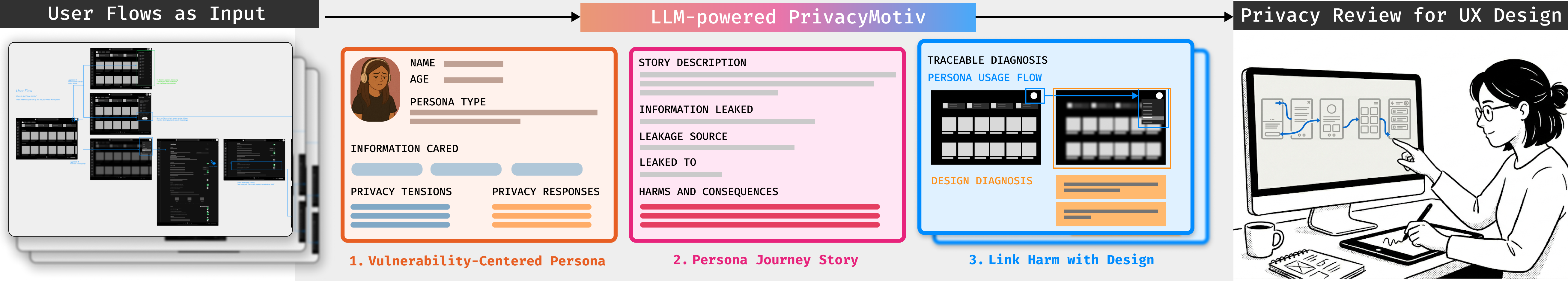}
  \caption{\textit{PrivacyMotiv} takes lo-fi user flows as input and generates three-tiered outputs: (1) vulnerability-centered personas, (2) persona journey stories, and (3) privacy harms and design linkage, to support privacy review in early-stage UX design.}
  \Description{A horizontal diagram illustrating the PrivacyMotiv workflow. On the left, a stack of lo-fi user flow diagrams serves as system input. An arrow points right into a pink-highlighted section labeled "LLM-powered PrivacyMotiv," which contains three color-coded output panels. The first panel, outlined in orange and labeled "Vulnerability-Centered Persona," shows structured fields including name, age, persona type, information cared about, privacy tensions, and privacy responses. The second panel, outlined in pink and labeled "Persona Journey Story," shows fields including story description, information leaked, leakage source, leaked to, and harms and consequences. The third panel, outlined in blue and labeled "Link Harm with Design," shows annotated lo-fi wireframe flows with persona usage flow highlighted in blue and design diagnosis highlighted in orange. A final arrow points to an illustration of a UX designer reviewing the outputs on a desktop screen, labeled "Privacy Review for UX Design.}
  \label{fig:teaser}
\end{teaserfigure}

\maketitle

\section{Introduction}

User Experience (UX) professionals typically follow structured frameworks like the Double Diamond~\cite{banathy2013designing} or Design Thinking~\cite{brown2008design} to ensure user-centricity. These workflows guide designers from \textit{Empathize} (understanding users), \textit{Define} (articulating needs), \textit{Ideate} (brainstorming solutions), to \textit{Prototype} and \textit{Test}~\cite{stickdorn_schneider_service_design_thinking_2011}. Yet, despite this foundational focus on empathizing with user needs, a critical ``empathy gap'' persists, particularly around user privacy~\cite{chenWen2025speculating}.

The Spotify ``Friend Activity'' feature serves as a compelling example, showing users what their friends are playing. Developed in response to over 7,000 user requests for social connection~\cite{spotifyCommunity,spotifySupportFriendActivity}, the feature addressed a genuine user desire but inadvertently enabled forms of location tracking and emotional surveillance~\cite{firm_alfalfa_6017_you_2024}. As one user review revealed: \textit{``After we broke up... I see her friend activity on Spotify... She only listens to music in the car, so if she is on Spotify, she is going somewhere''}~\cite{firm_alfalfa_6017_you_2024}. 
This case illustrates how a seemingly benign feature can be repurposed to harm individuals in vulnerable situations, such as between ex-partners, where one person can infer sensitive information about the other (e.g., location or routines) without their awareness, amplifying pre-existing power asymmetries through the app design. Such harms are not unique to Spotify; Citizen's neighborhood livestreaming~\cite{10.1145/3544548.3581258}, Venmo's public transaction feeds~\cite{notopoulos2021venmo}, and Snapchat's location-sharing~\cite{huie2023snapchat} reveal the same pattern: well-intentioned features can become instruments of harm when UX design ignores pre-existing social conditions and power asymmetries.

Synthesizing prior research on why UX practitioners rarely engage in privacy-aware design~\cite{wong_bringing_2019, lee_i_2024, zhang-kennedy_navigating_2024, li2018coconut} and their empathy deficit~\cite{marsden_stereotypes_2016, zhu_creating_2019, choi_proxona_2025}, we find that the issue is not simply missing tools, but low intrinsic motivation to embed privacy~\cite{chenWen2025speculating} in work. Framed by Self-Determination Theory (SDT)~\cite{ryan2000self}, three barriers undermine designers' motivation: a lack of \textit{connectedness}, as designers have limited knowledge of privacy harms and struggle to relate to affected users~\cite{shao_privacylens_2025, wong_bringing_2019, lee_i_2024}; a lack of \textit{competence}, since existing methodological formats (such as compliance checklists~\cite{matte_cookie_2020,nouwens_dark_patterns_after_gdpr_2020, grasl_dark_2021, gunawan_comparative_2021, gunawan_redress_2022}, static taxonomies~~\cite{mathur_what_2021,bosch_tales_2016, forbrukerradet_deceived_by_design_2018, forbrukerradet_every_step_you_take_2018}, and post-hoc evaluations~\cite{marsden_stereotypes_2016, turner_is_2011,Hertzum2003EvaluatorEffect}) rarely connect abstract principles to specific design problems or solutions~\cite{mildner_about_2023, nie_shadows_2024}; and a lack of \textit{autonomy}, as privacy incidents are viewed as a legal or executive decision, and the designer has no agency to intervene~\cite{zhang-kennedy_navigating_2024}. 

Addressing these gaps requires interventions beyond compliance audit; we need to extend designers' empathy toward users with vulnerabilities and restore their motivation to proactively engage in privacy-preserving practices. To bridge these deficits, we propose \textbf{\textit{PrivacyMotiv}}, a novel privacy review approach consisting of a three-tiered framework (\autoref{fig:teaser}). First, it generates vulnerability-centered personas (e.g., a gig worker or an abuse survivor) that help designers engage beyond generic user archetypes, challenging normative assumptions. Second, it transforms static user flows into contextualized persona journeys via LLM-generated narratives that simulate how specific design choices unfold into plausible harms. Finally, to ensure actionability, it provides traceable diagnoses by mapping these narrative harms to specific steps in the low-fidelity user flow. By connecting the ``who'' (user with vulnerabilities), the ``what'' (anticipated experience), and the ``where'' (UI fault), the system reframes privacy from a checklist task into a structured reflective practice. \textit{PrivacyMotiv} is designed for integration into early stages of existing UX workflows, operating on low-fidelity user flows so designers can identify and mitigate privacy issues before development begins.

We evaluated this approach through a within-subjects study with professional UX practitioners ($N=16$). Participants conducted privacy reviews for two app features using two different approaches: their own \textit{self-proposed approach} (representing the status quo of their professional practice) and \textit{PrivacyMotiv}. To mitigate order effects, we used a counterbalanced design with fictional features inspired by real-world designs that have led to privacy harms.

Our results show \textit{PrivacyMotiv} significantly outperformed the status quo practices in both subjective and objective measures. Participants reported higher levels of intrinsic motivation and empathy when using the tool. More importantly, this emotional engagement translated into tangible outcomes: participants identified 59\% more privacy issues and proposed 70\% more redesign solutions compared to their self-proposed baseline. Issues identified with \textit{PrivacyMotiv} were also more specific, leading to more concrete improvement proposals and broader coverage across Privacy by Design principles~\cite{cavoukian2009privacy}. Many participants shifted from a rigid, compliance-focused view of privacy (e.g., ``\textit{it’s basically about data permissions}'') to a contextual perspective that considered material consequences and harms to users and bystanders. Designers expressed optimism about adopting such lightweight, narrative-driven interventions, noting that they effectively augmented their professional judgment without disrupting their existing workflows.

This paper contributes to the usable privacy and design research by: 
(1) proposing a novel \textbf{privacy review approach} integrated into existing UX design workflows that shifts privacy assessment from a compliance task to a reflective practice, fostering empathy and intrinsic motivation; 
(2) operationalizing this approach through \textbf{\textit{PrivacyMotiv}, an LLM-powered system} grounded in vulnerability-centered personas and user flows that generates contextualized narratives of privacy risks; and 
(3) presenting \textbf{empirical findings} from a within-subjects study  with professional UX practitioners ($N=16$), showing that this narrative-driven approach significantly increases designers' empathy and motivation, outperforming current practices in identifying privacy harms and generating solutions. 
\section{Background and Related Work}

\subsection{Privacy Issues in UX Design: Beyond Malicious Intent to Unintended Harms}
Privacy violations in modern digital products extend beyond unauthorized data leakage, spanning institutional (platform-level misuse) and interpersonal dimensions (one user surveilling or harming another through app features), with the latter being particularly dependent on users' pre-existing social conditions and power asymmetries~\cite{nissenbaum2004privacy}. Contemporary systems have been found to trigger affective discomfort described as ``creepy'' or ``disconcerting''~\cite{seberger_still_2022, chenWen2025speculating} while impacting broader social values, including autonomy, physical safety, and trust~\cite{wong_eliciting_2017}. Citron and Solove's typology broadens this understanding, describing how privacy violations can lead to physical, economic, and reputational harms, as well as relationship damage~\cite{citron2022privacy}. Usable privacy research has also empirically documented in everyday digital interactions, including social surveillance, relationship harm, and loss of autonomy~\cite{10.1145/3469845, 238283}.

While early research at the intersection of privacy and UX focused heavily on deceptive design crafted to trick users~\cite{deceptive_design, bosch_tales_2016}, recent practitioner-centered research highlights a more complex reality: designers often struggle to navigate ethical ambiguity in real-world contexts~\cite{zhang-kennedy_navigating_2024}. Many privacy harms do not stem from malice but from ``privacy gray patterns''~\cite{zhang-kennedy_navigating_2024}, designs that serve legitimate functional purposes and follow industry norms yet inadvertently enable privacy violations. Real-world incidents, such as the ones in Spotify's ``Friend Activity''~\cite{firm_alfalfa_6017_you_2024, nickc0sta_spotify_2021, redditSpotifyComplaints, techcrunchSpotifyCommunity}, exemplify that even well-intentioned or widely adopted designs can become instruments of harm when usage contexts shift.

Despite this growing awareness, existing support methods remain misaligned with design practice. Methods such as compliance checklists~\cite{matte_cookie_2020, nouwens_dark_patterns_after_gdpr_2020} or engineering frameworks like LINDDUN~\cite{wuyts2020linddun} tend to treat privacy as a legal obligation rather than a user experience quality, making it hard for designers to anticipate how UI/UX decisions can lead to real-world harms~\cite{10.1145/3180155.3182531, zhang-kennedy_navigating_2024, li2018coconut, prybylo2024evaluating}. Thus, there is a critical need of support that goes beyond compliance to surface unintended consequences within existing design artifacts.



\subsection{Gaps Between Privacy Principles and UX Practice}\label{sec:motivation-barrier}


The third wave of security and privacy, inclusive security and privacy, emphasizes mechanisms that accommodate diverse characteristics, abilities, needs, and values, ensuring underserved populations can protect their privacy and security \cite{10.1145/3171533.3171538}. These vulnerable groups often face distinct, disproportionate privacy harms overlooked in mainstream design and policy~\cite{mcdonald2022privacy, sannon2022privacy}. This shift has led to a growing body of research in the HCI and usable security and privacy communities that investigates the unique challenges faced by marginalized and vulnerable populations. These studies span a wide range of user groups, including individuals in resource-constrained regions~\cite{vashistha2018examining}, sex-trafficking survivors~\cite{gautam2020usable}, minoritised ethnic communities~\cite{quyoum2025minoritised}, refugees facing online toxic content~\cite{arunasalam2024understanding}, survivors of intimate partner abuse~\cite{matthews2017stories}, people with disabilities~\cite{zezulak2023sok}, older adults~\cite{knight2023privacy, aly2024tailoring, das2024design}, and more. A recent review by \citet{sannon2022privacy} introduces the ``Privacy Responses and Costs'' framework, highlighting the unique tensions these groups face. These risks challenge normative assumptions about what privacy is and who it is for, underscoring the need for approaches that foreground the diverse needs of at-risk users from the outset of design~\cite{mcdonald2020privacy, 10.1145/3469845}.

Despite this wealth of theoretical and empirical knowledge, a significant gap remains in transferring them into design practice. Prior work suggests UX practitioners often underprioritize privacy-aware design due to limited \textit{intrinsic motivation} to embed privacy into their work~\cite{chenWen2025speculating, marsden_stereotypes_2016}.
Designers often miss how everyday UI flows can be weaponized or cause downstream privacy leaks~\cite{sannon2022privacy, shao_privacylens_2025, chen2025engagementprolongingdesignsteensencounter}. Abstract privacy principles don’t translate cleanly into UI decisions, making risks hard to spot in practice~\cite{wong_bringing_2019, li2018coconut, bongard-blanchy_i_2021}. Privacy is then treated as fixed compliance or industry norm, leaving designers feeling unable to push for change~\cite{zhang-kennedy_navigating_2024, prybylo2024evaluating}.
Addressing empathy and motivation deficits requires early interventions that motivate designers by connecting abstract privacy risks to the situated consequences of users with vulnerabilities, fostering empathy during the early stages of the design process.

\subsection{Empathy Building: From Static Personas to Contextual Narratives}
To bridge the gap between designer intent and user reality, empathy has long been positioned as a cornerstone of user-centered practice~\cite{kouprie_sleeswijk_visser_empathy_design_2009, hess_manifestation_2016}. The most common method, the user persona, aims to help designers ``step into the user's shoes.'' However, traditional personas often rely on ``imagined users'' that simplify complex realities or rely on stereotypes, failing to represent perspectives beyond the designer's immediate context~\cite{marsden_stereotypes_2016, turner_is_2011}. Furthermore, empathy is predominantly emphasized during early discovery phase, with limited application in later design review stages.

To address the limitations of static personas, researchers have introduced speculative narrative and scenario-based methods that surface concerns across wide-ranging contexts. Examples include \citet{wong_eliciting_2017}'s design workbooks for privacy-invasive technologies, \citet{pierce_expanding_2015}'s metaphors for surveillance, and \citet{h_tan_monitoring_2022}'s in-home speculative probes. Though promising for building vulnerability-aware, context-rich understanding, these methods often function as standalone workshops or high-level exercises. Our work seeks to operationalize this narrative approach into everyday UX practice by anchoring it into specific design artifacts (user flows) to trigger empathy not just for the user, but for the specific moment when harm occurs.

\subsection{LLM-Powered Design Support}
LLMs present new opportunities for scaling empathetic design, supported by recent research demonstrating that their social situational judgment capabilities match or even outperform humans~\cite{mittelstadt2024large}. While AI-generated personas are often more informative and consistent than human-crafted ones, they risk reflecting stereotypes, revealing both the promise and the need for greater fidelity and representation~\cite{lazik2025impostor}. Shin et al.~\cite{shin2024understanding} find human-AI collaboration yields more representative and empathy-evoking personas. Distinct strategies have emerged to balance this efficiency with variety and fidelity~\cite{schuller2024persona, shin2024understanding}: PrivacyLens~\cite{shao_privacylens_2025} evolves privacy-sensitive seeds into persona vignettes; Persona-L~\cite{sun2025personal} utilizes a chat-based approach to define complex user needs; and \citet{mei_geneymap_2025} leverage LLMs for persona journey mapping.

In broader design support, LLMs are increasingly used to augment the design lifecycle. \citet{ahmed2025role} reviewed the role of conversational agents in UI/UX, systems like GenieWizard~\cite{yang2025geniewizard} and Canvil~\cite{feng2025canvil} utilize LLMs to analyze design skeletons or facilitate iterative prototyping. \citet{duan2023towards} developed an LLM-based heuristic evaluation plugin for Figma. A closely related work, Farsight~\cite{wang2024farsight}, uses AI to help designers identify potential harms in AI applications. 
However, relatively little research has explored using LLMs to tackle designers' empathy and motivation deficits in addressing the unintended user behaviors in UX design. Our approach, \textit{PrivacyMotiv}, addresses this gap by focusing on privacy review---an area where these pain points are especially acute---and by leveraging a theoretically grounded, LLM-based pipeline to ensure both generation quality and scalability for integration into practical design workflows.
\section{PrivacyMotiv Design and Implementation}

\begin{figure*}[t]
    \centering
    \includegraphics[width=\textwidth]{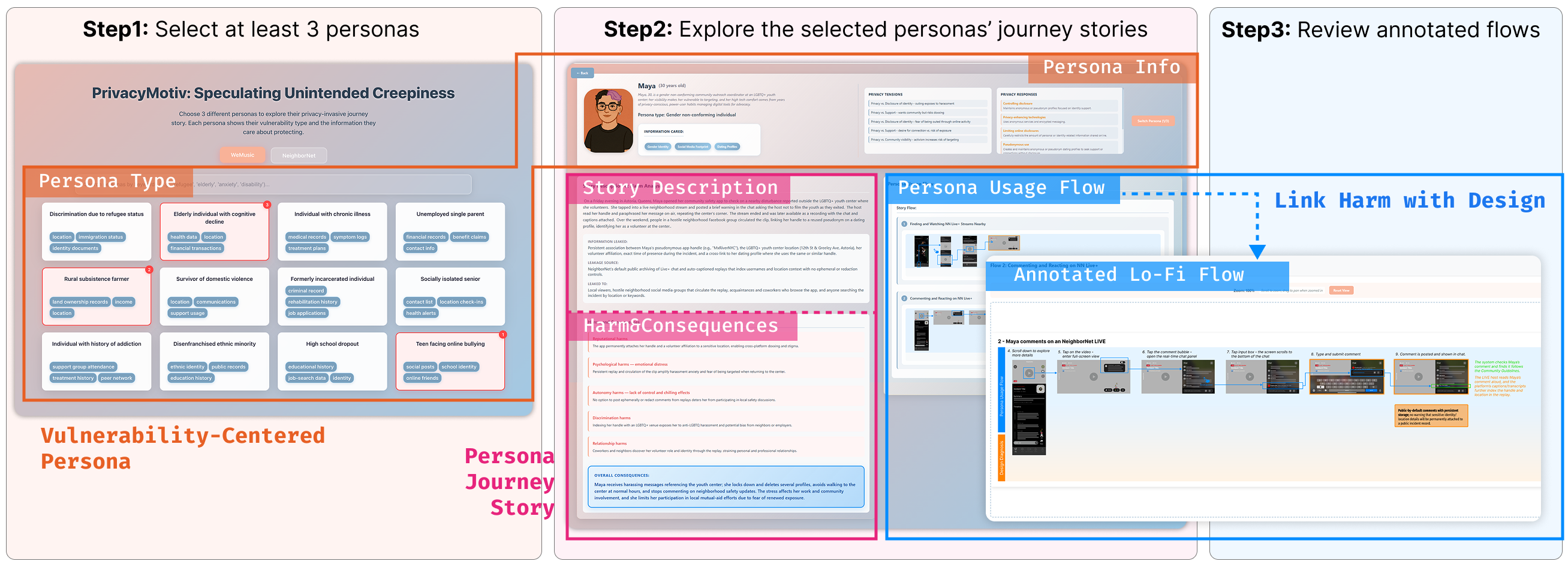}
 \caption{\textbf{\textit{PrivacyMotiv} System Overview.} The end-to-end workflow consists of three steps: (1) Persona Selection: users choose from vulnerability-centered personas; (2) Personal Journey Story: the system generates narrative-style user journeys demonstrating how privacy harms emerge from app interactions; and (3) Harm-Design Linkage: annotated low-fidelity user flows let designers trace privacy harm triggers, with color-coded annotations marking user actions (blue) and design flaws (orange).}
  \Description{The PrivacyMotiv system interface divided into three panels. Left panel (Step 1): A persona selection grid displaying vulnerability-centered persona types organized by category, such as teens facing online bullying and domestic violence survivors. Center panel (Step 2): A narrative persona journey page showing a selected persona's profile, a story description of a privacy incident, and a harms and consequences section. Right panel (Step 3): Annotated low-fidelity wireframe user flows with color-coded callouts — blue for user actions and orange for design flaws — enabling designers to trace where privacy harms are triggered in the interface.}
    \label{fig:privacyMotiv_overview}
\end{figure*}

To address privacy issues in UX design grounded in situated user experiences, we developed \textbf{\textit{PrivacyMotiv}}, a UX review assistant tool that shifts privacy review from a retrospective compliance check to a proactive activity embedded in early-stage design workflows (e.g., before hi-fi prototyping and engineering). 

Informed by the empathy and motivation deficits detailed in Section~\ref{sec:motivation-barrier}, we define three system design goals that draw on Self-Determination Theory (SDT)~\cite{ryan2000self} which identifies three core psychological needs supporting intrinsic motivation:  \textit{connectedness} (feeling relatable to others), \textit{competence} (feeling capable of effective action), and \textit{autonomy} (feeling in control over one's choices). Together, these goals aim to restore designers’ intrinsic motivation and empathetic engagement with privacy.

\begin{description}
    \item[D1:] \hypertarget{D1}{\textit{\textbf{Center on the diverse privacy needs of vulnerable populations.}}}
\end{description}
Designers often rely on ``imagined users'' resembling themselves, leading to blind spots around risks faced by marginalized groups. To foster \textit{connectedness}, the system generates personas defined by specific vulnerabilities rather than generic user archetypes, ensuring these perspectives are foregrounded in design critique.

\begin{description}
    \item[D2:] \hypertarget{D2}{\textit{\textbf{Engage with privacy through contextualized, speculative persona journeys.}}} 
\end{description}
To address the lack of empathetic engagement with privacy issues that emerge during user interactions, the system uses LLM-generated narratives to simulate how specific design decisions impact the experiences of users with certain vulnerabilities, enhancing \textit{connectedness} while helping designers foresee plausible risks and build \textit{competence}.

\begin{description}
    \item[D3:] \hypertarget{D3}{\textit{\textbf{Make the relationship between design choices and privacy harms visible and traceable.}}} 
\end{description}
Empathy alone is insufficient without actionable pathways. Designers often view privacy as a legal issue, or feel powerless to challenge industry norms and business goals without concrete evidence (\textit{autonomy}). They also report less confidence in proposing alternatives when unable to identify the specific design element at fault (\textit{competence})~\cite{wong_bringing_2019, zhang-kennedy_navigating_2024, chenWen2025speculating}. To support this, the system traces each harm back to specific UI elements and flow steps within the designer’s sphere of influence, empowering them to diagnose and address privacy risks through concrete, justifiable interventions.

\subsection{PrivacyMotiv System Overview}

The target users of \textit{PrivacyMotiv} are UX design practitioners.
\textit{PrivacyMotiv} analyzes UX user flows through vulnerability-centered personas and generates contextualized persona journey stories that reveal potential harms for users in vulnerable situations (\autoref{fig:privacyMotiv_overview}).
The system operates in three stages, each aligned with one or more of our core design goals (\textbf{\hyperlink{D1}{D1}}, \textbf{\hyperlink{D2}{D2}}, \textbf{\hyperlink{D3}{D3}}). 

First, it displays pre-generated vulnerability-centered personas (\autoref{subsec:persona_prompt}) for designers to select. These personas capture established vulnerability contexts grounded in domain-specific literature, which are constructed for the general privacy harm exploration and demonstration purposes, rather than customized for a specific application.
Second, it generates contextualized persona journey narratives by prompting an LLM with a selected persona paired with researcher-constructed combinations of user flows, strictly without introducing new or hallucinated flows (\autoref{subsec:story}). These narratives form the dynamic layer of the system: they vary with different combinations of the input flows, and reveal the situated harms experienced by persona user with particular vulnerabilities.
Finally, it presents visual storyboards with annotated flows (\autoref{subsec:visual}), which communicate insights via low-fidelity (lo-fi) diagrams, a format closely aligned with designers’ review practices.
We further illustrate how the system can be used in practice through an example usage scenario (\autoref{subsec:usage_scenario}).

\subsection{Vulnerability-Centered Persona Generation}
\label{subsec:persona_prompt} 

\begin{figure*}[t]
    \centering
    \includegraphics[width=\textwidth]{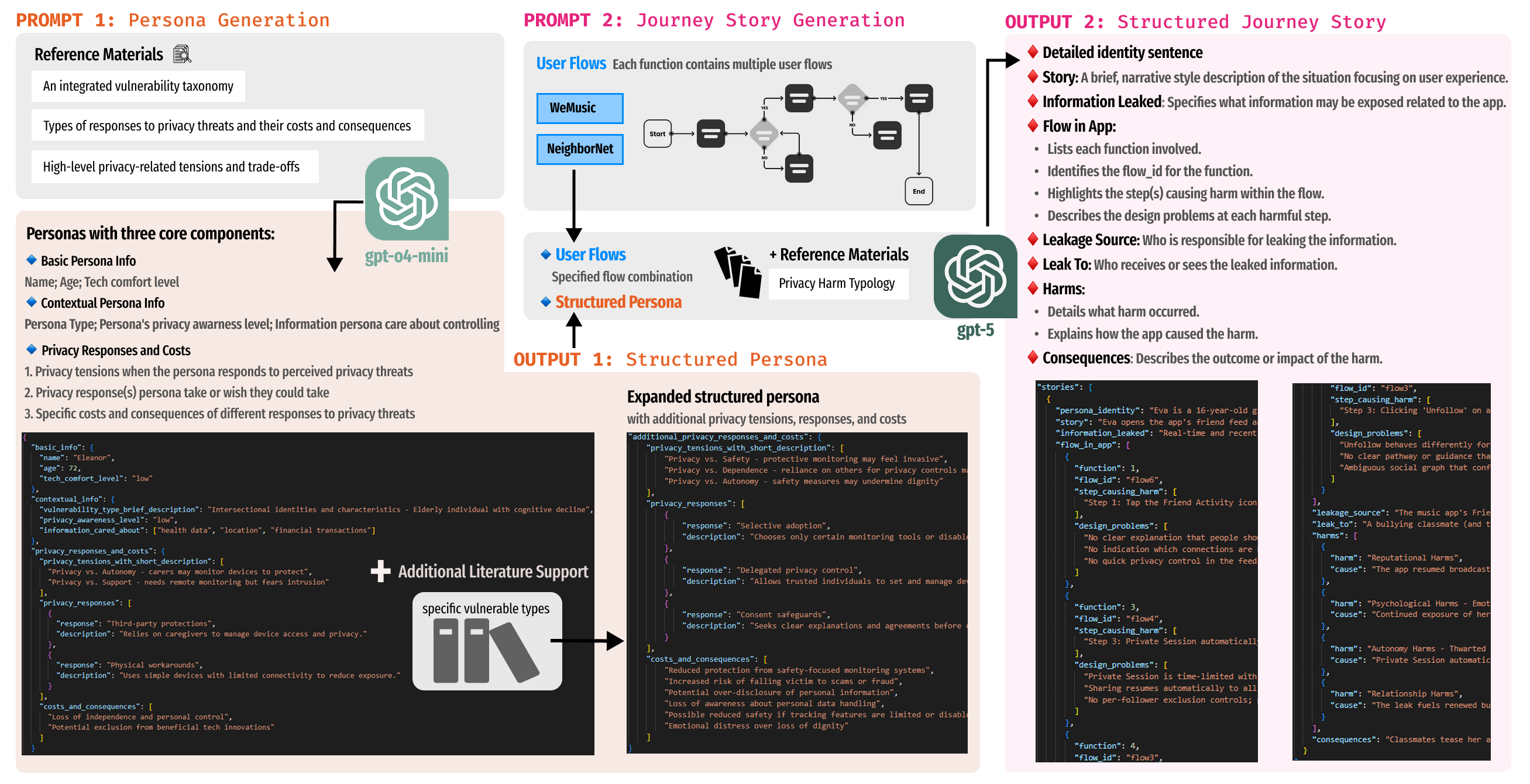}
    \caption{
    The two-stage generation pipeline of \textit{PrivacyMotiv}. \textcolor{orangeOrange}{Prompt1} generates initial structured personas (\textcolor{orangeOrange}{Output1}) grounded in a vulnerability taxonomy and privacy marginalization literature (Appendix~\ref{appendix:Input_1}). \textcolor{pinkPink}{Prompt2} combines these personas with user flows and a Privacy Harms Typology to generate structured journey stories (\textcolor{pinkPink}{Output2}) (Appendix~\ref{appendix:Input_2}). }
    \Description{A diagram illustrating the two-stage generation pipeline of PrivacyMotiv, arranged left to right. On the left, Prompt1 shows reference materials, including an integrated vulnerability taxonomy and privacy and marginalization research key constructs---fed into gpt-o4-mini to generate Output1: initial structured personas with three core components: basic persona info, contextual persona info, and privacy responses and costs. These personas are then enriched with additional literature support specific to vulnerable types, resulting in expanded structured personas. On the right, Prompt2 combines these expanded personas with user flows and a Privacy Harms Typology reference, fed into gpt-5 to produce Output2: structured journey stories. Each journey story includes components such as a detailed identity sentence, story, information leaked, flow in app, leakage source, harms, and consequences. Together, Output1 and Output2 form the PrivacyMotiv system.}
    \label{fig:generation_pipeline}
\end{figure*}

Vulnerable populations face distinct and often disproportionate privacy harms that are frequently overlooked in mainstream design processes ~\cite{mcdonald2022privacy, sannon2022privacy}. To help designers recognize and center these needs, \textit{PrivacyMotiv} incorporates vulnerability-centered personas to promote empathetic understanding (\textbf{\hyperlink{D2}{D2}}) and support inclusion of these users (\textbf{\hyperlink{D1}{D1}}).

\paragraph{\textbf{Persona schema}}
Each \textit{persona} is a structured profile composed of three core components: (1)~\textit{basic demographic information}, including the individual's name, age, and technology comfort level; (2)~\textit{contextual vulnerability details and privacy concerns}, incorporating the persona type, the user’s privacy awareness level, and the types of personal information they seek to protect; (3)~\textit{privacy responses and consequences}, capturing the specific privacy tensions the user experiences, their behavioral responses, and the associated costs or consequences. This structured representation ensures that each persona is grounded in plausible, situated user experiences. It also surfaces design-relevant insights into how privacy harms manifest in the lives of users with vulnerabilities.

The \textit{persona type}, included in the contextual vulnerability component, serves as a representative user category reflecting structural, social, and situational factors contributing to privacy risks (key vulnerability dimensions as outlined in~\autoref{appendix:vulnerability-taxonomy}) . Each persona type represents a user group, such as a person with physical disability, gender non-conforming individual, or elderly individual with cognitive decline. To reduce the risk of essentialization, each persona is anchored to a single salient vulnerability dimension rather than compounding multiple axes of identity~\cite{tang2025beyond, turner_is_2011}.

\paragraph{\textbf{Persona generation}}
As discussed in \autoref{sec:motivation-barrier}, empirical insights into privacy issues faced by vulnerable users exist but are scattered across a large volume of papers.
To systematize this knowledge, we curated an integrated vulnerability taxonomy by synthesizing prior research defining and categorizing vulnerable populations~\cite{limante2022definition, Rukmana2014, numans2021vulnerable}. 
By synthesizing across sources, we aimed to (1) capture a broad and diverse set of vulnerability categories, and (2) clarify overlapping concepts without distorting their original meanings. Originating from legal, sociological, and insider perspectives, these sources vary in emphasis and terminology but often overlap in substance. For instance, definitions may frame vulnerability through legal exclusion, social stigmatization, or impaired self-reliance, yet they frequently converge on common groups (e.g., people with disabilities, the elderly, migrants) and conditions (e.g., reduced resource access, increased harm exposure, limited social participation). We clarified such overlaps by grouping them under distinct dimensions while preserving the original intent.
This synthesis process resulted in an integrated vulnerability taxonomy (\autoref{appendix:vulnerability-taxonomy}) that preserves the nuance of the original sources while supporting downstream persona generation for privacy design analysis.

We then generate personas by extracting detailed privacy issues from research papers that are associated with specific vulnerability dimensions in the taxonomy.
\citet{sannon2022privacy} present a literature review that analyzes 88 papers published between 2010 and 2020 on privacy and marginalization and proposes constructs to capture key details for understanding privacy and marginalization, including ``responses to privacy threats,'' ``costs and consequences,'' and ``privacy-related tensions and trade-offs.'' We incorporate these constructs into our personas’ structure. As a starting point, we categorize evidence cited in \citet{sannon2022privacy} according to the vulnerability types in our taxonomy to craft an initial set of personas for users with vulnerabilities.

The generation is initially performed by \verb|gpt-o4-mini|, which is well suited for processing unstructured data and reformatting it into structured profiles at scale, followed by researchers’ verification. From the initial set of 30 generated personas, we selected 20 for inclusion in the study (~\autoref{appendix:persona_types}). The selection process removed those that were redundant, insufficiently distinct, or whose persona type vulnerabilities had limited relevance to the contexts of the experimental application. 

To enrich the personas, we further conducted focused literature searches to find additional empirical studies, prioritizing work in usable privacy and HCI domains, to ensure topical relevance and contextual fit. 
These sources enriched the \textit{privacy responses and consequences} component by supplementing LLM-generated content with empirically grounded insights. LLMs extracted and reformatted relevant information, which researchers validated through cross-referencing and manual refinement to ensure fidelity and representational accuracy (see~\autoref{fig:generation_pipeline}, \textcolor{orangeOrange}{Output1}). We treat these personas as empirically informed starting points for design reflection, rather than authoritative representations of any group's lived experience, acknowledging that centering vulnerability dimensions is a deliberate choice that risks essentialization.

\subsection{Speculative Persona Journey Story Generation: Contextualizing Privacy Tensions and Harms}
\label{subsec:story}

\subsubsection{User Flows as Input}

\begin{figure*}[htbp]
    \centering
    \includegraphics[width=\textwidth]{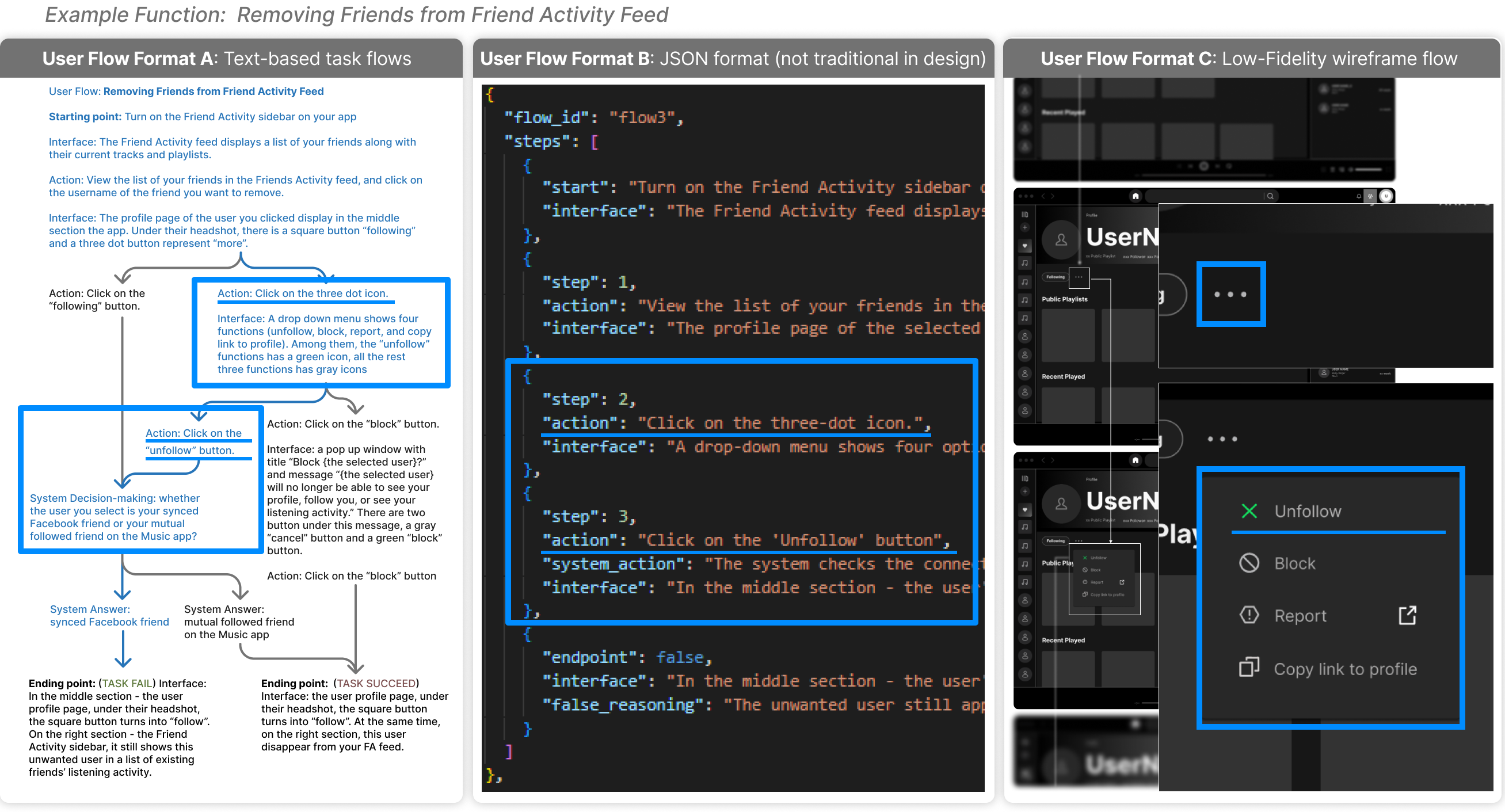}
    \caption{Three representations of the same user flow in the function ``Removing Friends from Friend Activity Feed'', each serving a different role in the \textit{PrivacyMotiv} system.  \textbf{(A)} A traditional text-based task flow drawn in Figma, describing user actions and interface states in text. \textbf{(B)} A JSON-formatted user flow mapped from the text-based version to support system parsing and privacy narrative generation; not visible to designers. \textbf{(C)} A low-fidelity wireframe version of the same flow, used as the main review artifact for designers.}
    \Description{An overview showing three representations of the same user flow for the function Removing Friends from Friend Activity Feed. Panel A shows a traditional text-based task flow drawn in Figma, describing user actions and interface states in text. Panel B shows a JSON-formatted version of the same flow, structured for system parsing and privacy narrative generation, which is not visible to designers. Panel C shows a low-fidelity wireframe version of the same flow, used as the main review artifact presented to designers during the study.}
    \label{fig:user_flow_format}
\end{figure*}

To ground the speculative persona journey in concrete app design, we use user flows, a commonly used artifact in real-world UX practices that representing detailed interaction logic. In our framework, we distinguish between features, functions, and user flows. A feature is a higher-level capability in the app (e.g., WeMusic’s Friend Activity). Each feature consists of multiple functions, which are specific user tasks supported within that feature (e.g., viewing friend activity, adding friends via Facebook connection, starting a private session, or removing friends). Each function can be decomposed into multiple user flows, where a user flow represents a possible linear path a user can execute to accomplish the function.

User flows in design practice vary in fidelity and modality, ranging from diagrams (e.g.,~\autoref{fig:user_flow_format}A) to lo-fi wireframes (\autoref{fig:user_flow_format}C) and hi-fi prototypes, but they share a common underlying structure. To enable system-level interpretation, we introduce a structured intermediate JSON representation (\autoref{fig:user_flow_format}B) that mirrors these elements, including a unique \verb|flow_id| and an ordered list of \verb|steps|. Each step includes fields such as the \verb|step| number, \verb|action| (what the user does), \verb|interface| (what is shown on screen), and optionally \verb|system_action| (system response). The flow concludes with an \verb|endpoint| flag and a \verb|true_reasoning| or \verb|false_reasoning| description of the outcome.

In this work, user flows were manually prepared by researchers based on fictional apps reproduced from real-world apps and features (detailed in~\autoref{subsubsec:fictional_apps_and_materials}). These flows were presented as lo-fi wireframes, mirroring real-world UX review practices where flows are prepared early in the design process and reviewed by external designers who are often unfamiliar with the app. For system input, researchers mapped these wireframe flows into the JSON format described above; while participants interacted with lo-fi wireframes, the underlying system input was the corresponding JSON flows with identical content. LLMs are scoped specifically to analyzing these input flows and generating narrative persona journeys for privacy contextualization, strictly without constructing, converting, or hallucinating new flows.

%

\subsubsection{Harm-Informed Persona Journey Story and Design Diagnosis}
\label{subsubsec:user_journey_generation}
We prompt a state-of-the-art reasoning model \verb|gpt-5| to generate structured persona journey stories and associated design diagnoses (\autoref{appendix:Input_2}). Given a persona and a designated sequence of app functions, each comprising multiple user flows, the model generates a user journey grounded in the actual design, together with a design diagnosis that identifies privacy-violating elements.

Narrative generation is guided by Citron and Solove’s typology~\cite{citron2022privacy} to identify situations that may cause privacy harms. First, the system enriches each persona with a concise identity sentence describing their demographic attributes, role, and how these connect to their vulnerability. It also characterizes the persona’s comfort level with technology, rooted in everyday experiences.

Next, for each function in the assigned sequence, the model is prompted to select a single user flow that aligns with the evolving context of the story. This structured combination of persona and function-flow level interaction serves as the foundation for the scenario-based user journey narratives, simulating how the persona might realistically engage with the app. Each resulting story reflects the persona's original privacy concerns and responses while demonstrating how specific interface elements and design decisions could lead to potential privacy harms (\textbf{\hyperlink{D3}{D3}}). See~\autoref{appendix:example-story} for a complete example of a persona journey narrative.

Generated outputs were carefully reviewed and excluded if LLM introduced app features absent from the actual flow, generated actions the interface could not support, deviated from the persona’s attributes and privacy concerns, or solely focused on outcomes (e.g., a data leak) without explaining the UI mechanisms that led to those harms. This filtering ensures outputs quality, analytical clarity, and traceability for designers to understand privacy risks to specific design choices.

The final output format is a structured narrative annotated with specific fields (\autoref{fig:generation_pipeline}, \textcolor{pinkPink}{Output2}): the user's story, potentially leaked sensitive information, the specific steps within each flow causing leakage, UI-level issues at problematic steps, and resulting harms and consequences. By emphasizing the unfolding process, this format shifts focus from abstract policy-level concerns to human-centered, design-level reflections on privacy.

\begin{figure*}[htbp]
    \centering
    \includegraphics[width=0.85\textwidth]{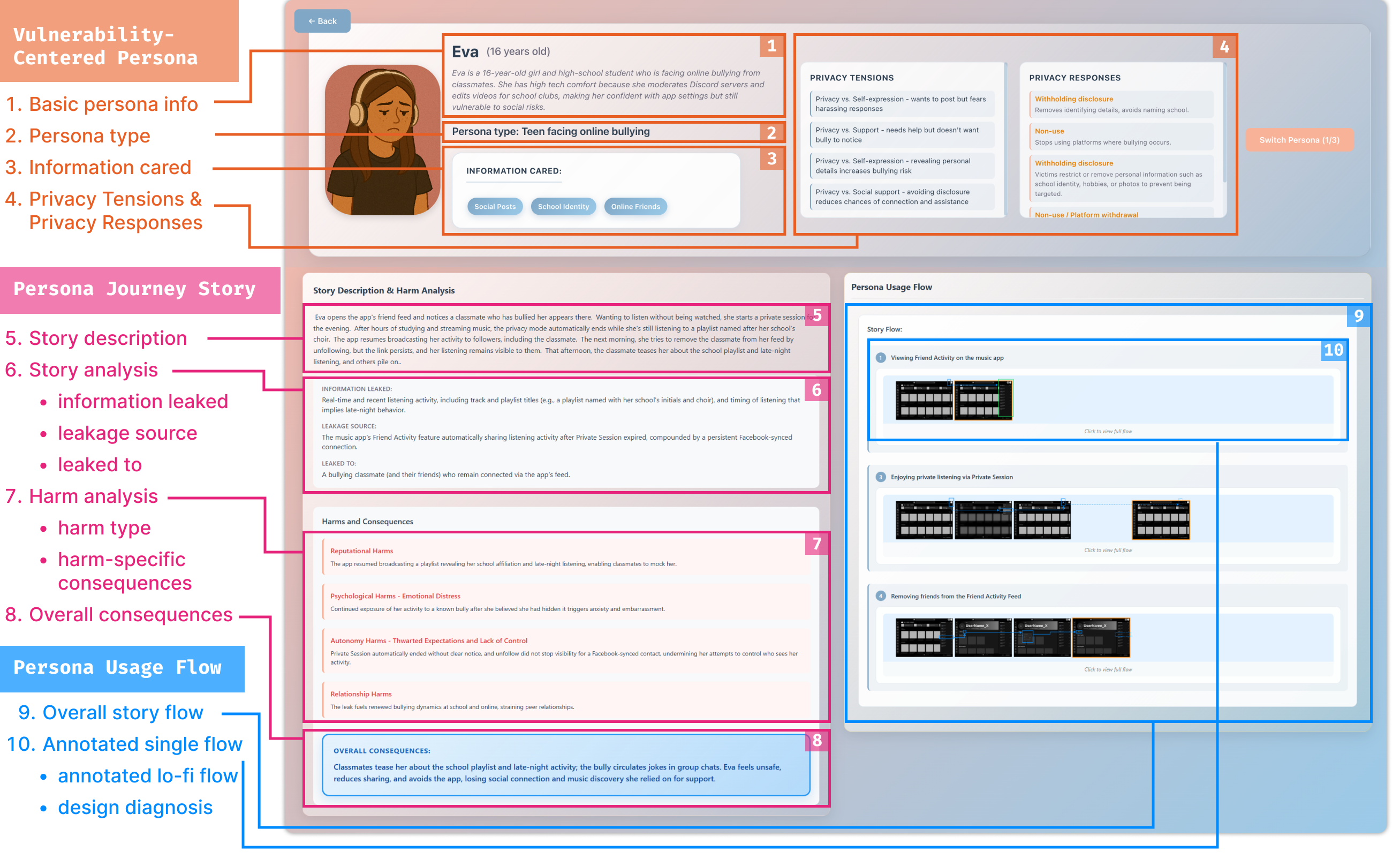}
    \caption{\textbf{\textit{PrivacyMotiv}’s main page overview (using Eva as an example)}. The interface aggregates three core sections to contextualize privacy harms: (1) \textcolor{orangeOrange}{Vulnerability-Centered Persona} (top bar), detailing the persona’s specific privacy issues; (2) \textcolor{pinkPink}{Persona Journey Story} (bottom left), describing the specific privacy incident and resulting consequences; and (3) \textcolor{blueBlue}{Persona Usage Flows} (bottom right), mapping the narrative harms directly to specific steps in the low-fidelity wireframes. For a detailed demonstration of Eva’s content, see~\autoref{appendix:example-story}.}
    \Description{An overview of PrivacyMotiv’s main interface using Eva, a teen facing online bullying, as an example persona, divided into three color-coded sections. The top bar highlighted in orange shows the Vulnerability-Centered Persona panel, displaying Eva’s basic persona info, persona type, information cared about, privacy tensions and privacy responses. The bottom left panel highlighted in pink shows the Persona Journey Story, presenting a story description and harm analysis including information leaked, leakage source, leaked to, harm types with harm-specific consequences, and overall consequences. The bottom right panel highlighted in blue shows the Persona Usage Flow, displaying an overall story flow and annotated single flows with lo-fi wireframes and design diagnostics that map narrative harms to specific interaction steps.}
    \label{fig:Eva_overview}
\end{figure*}

\subsection{UX of PrivacyMotiv: Making Design Consequences Visible and Traceable}
\label{subsec:visual}

\begin{figure*}[htbp]
    \centering
    \includegraphics[width=0.85\textwidth]{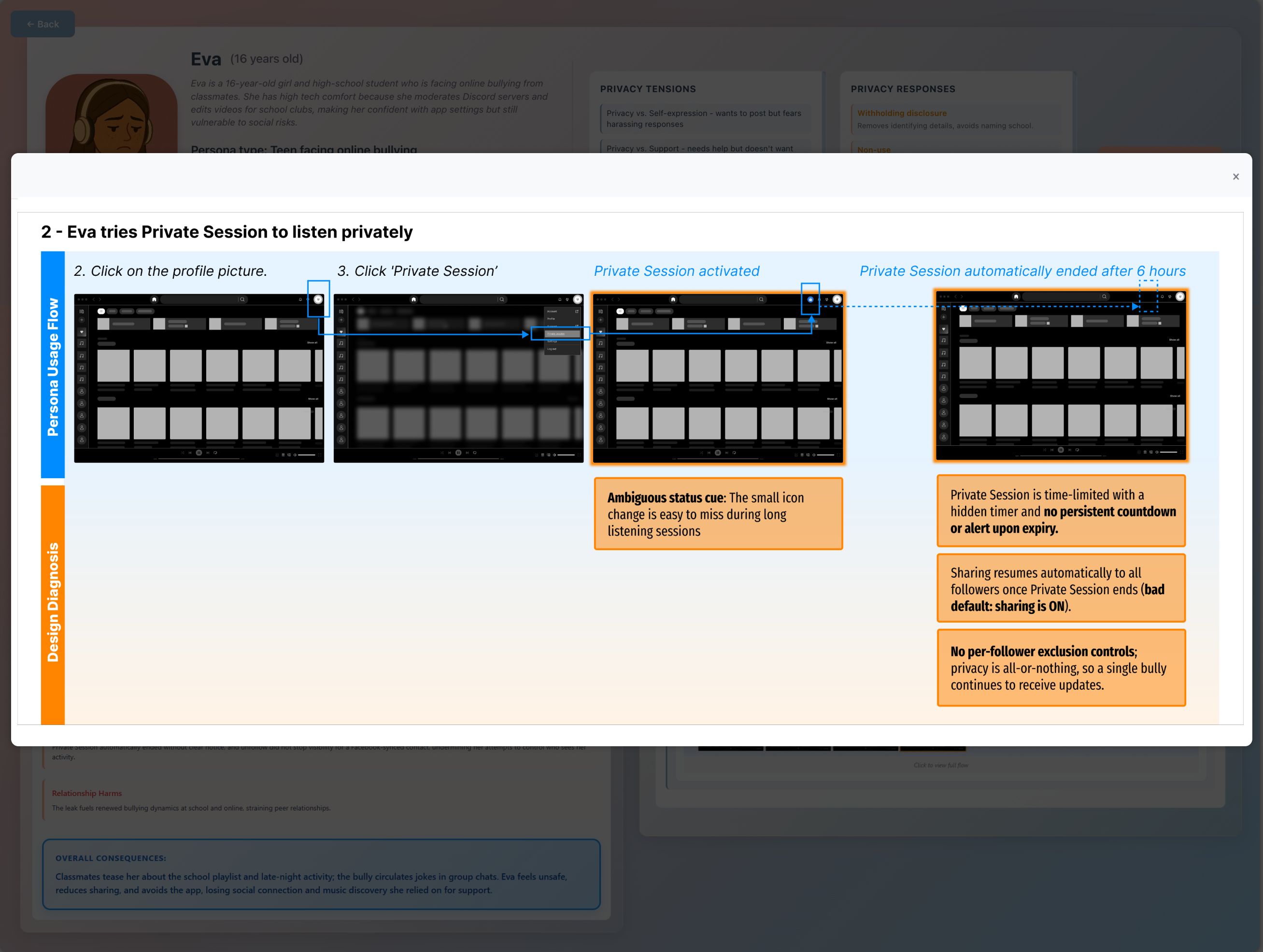}
    \caption{\textbf{Design Flaws Triggering Privacy Harms in Eva's Case.} This annotated user flow, presented in a low-fidelity wireframe format, illustrates how specific UI designs and default settings contributed to privacy harms for Eva.}
    \Description{An annotated low-fidelity wireframe user flow illustrating how specific UI design decisions and default behaviors in the Private Session feature contribute to privacy harms for Eva, one of the vulnerability-centered personas. The flow shows a sequence of wireframe screens with color-coded annotations tracing the specific interaction points where privacy violations occur and highlighting how the app's default settings and interface choices expose Eva to unintended surveillance and harm.}
    \label{fig:Eva_flow}
\end{figure*}

\subsubsection{Persona Selection and Analysis Results Presentation}

Upon launching the tool, designers are allowed to choose from a set of \textit{vulnerability-centered personas}, respecting their agency while encouraging greater engagement and empathy toward populations most relevant to the tested apps or their own personal experience (\textbf{\hyperlink{D2}{D2}}).
Each persona preview includes the persona type (e.g., refugees, older adults, and gender non-conforming individuals) and types of personal information they seek to protect, such as health data, location, and financial status.

On the following page, \textit{PrivacyMotiv} presents analysis results tailored to the selected persona (see~\autoref{fig:Eva_overview}). The detailed persona is foregrounded because designers typically begin by clarifying who the user is, establishing the human context necessary for empathetic engagement. The speculative \textit{persona user journey} with privacy harm analysis and \textit{persona usage flow} with lo-fi design diagnoses are then presented.

\subsubsection{Annotated Lo-fi User Flows for Linking Privacy Harms With Design}

A core mechanism of \textit{PrivacyMotiv} is linking privacy harms to specific interaction points through annotated lo-fi user flow storyboards. Each displayed wireframe flow corresponds to the input flows were selected by the LLM during narrative generation. Researchers then manually mapped the LLM-generated annotations onto the wireframe UIs, enabling designers to better trace where privacy violations occur (\textbf{\hyperlink{D3}{D3}}). These annotations, generated as part of the user journey story output (\autoref{subsubsec:user_journey_generation}), identify specific interface elements, default settings, and interaction patterns that contribute to harm, directly signaling to designers how privacy can be gradually violated during seemingly normal usage.


Annotations highlight problematic design decisions through color-coded callouts that identify specific interface flaws and their privacy implications: user actions (blue) and design flaws (orange). This visual mapping enables designers to see precisely how interface choices lead to privacy consequences. The system also provides zoom functionality to allow designers to examine interface details closely, and supports pan/drag navigation for comprehensive flow exploration. For example, in Eva's case (\autoref{appendix:example-story}), the flow titled \textit{``Enjoying private listening via Private Session''} reveals three design flaws: (1) the hidden session timer with no persistent countdown or alert upon expiry, (2) default resumption of sharing without user consent after the session expires, and (3) the lack of per-follower privacy controls, enforcing an all-or-nothing sharing model (see~\autoref{fig:Eva_flow}). By linking each harm to specific steps and interface choices within each flow, the system emphasizes that privacy violations may arise from default settings and overlooked user contexts, not just malicious intent.

\subsection{Example Usage Scenario}
\label{subsec:usage_scenario}
To illustrate how the system can be used in practice, we present the following usage scenario.

Alice, a professional UX designer, is tasked with conducting a privacy review of the early-stage design for a new social feature to be launched in the music app’s next version.
In prior workflows, such reviews typically happened internally within the UX team, or in cross-functional meetings with product managers, researchers, and engineers. These sessions often focused on usability, aesthetics, and feature completeness, with privacy considerations only occasionally mentioned. Without a structured approach to considering privacy from the user’s perspective, Alice finds it difficult to predict how real users might feel exposed or harmed by a design, especially for users whose privacy needs differ significantly from her own. While some privacy frameworks and guidelines are available, they are often too vague, legalistic, or difficult to apply in the early stages of prototyping. As someone without prior experience in privacy, she decides to try out \textit{PrivacyMotiv} to help with this task.

Using \textit{PrivacyMotiv}, Alice begins by selecting from a set of vulnerability-centered personas. She selects Eva, a 16-year-old high school student whose classmates’ online bullying makes her vulnerable to social risks. The system presents a narrative harm scenario grounded in the early-stage design of the same social feature Alice is reviewing. In the scenario, Eva starts a private listening session, but it expires while she is still listening to a school-choir playlist. Her listening activity is then shown in the app’s feed, prompting renewed teasing from the classmate and their friends. Uncommunicated session expiration and implicit visibility changes can inadvertently compromise user privacy and lead to harm.

Alice then explores several lo-fi user flows of the design under review on the right side of the same screen, consisting of a sequence of interface screens and system responses. The selected flows correspond  to the narrative harm scenario constructed by \textit{PrivacyMotiv}, which demonstrates how the app feature might function in practice, grounding the story in specific design elements. \textit{PrivacyMotiv} annotates the lo-fi user flows to help Alice trace how specific interactions, UI components, and system behaviors contributed to the harms in the scenario. For example, annotations highlight where limited user control may result in undesired visibility, such as the absence of quick privacy control in the feed to immediately restrict what specific people can see. By connecting the narrative to its underlying design flow, \textit{PrivacyMotiv} enables Alice to reason about privacy risks at the level of specific design decisions, surfacing concerns she might have overlooked in a traditional design review and fostering greater empathy for the affected users.

\subsection{System Scope and Implementation}
\label{subsec:system_scope_and_implementation}

\textit{PrivacyMotiv}\footnote{\url{https://peach-research-lab.github.io/PrivacyMotiv/}} is implemented as a web app built with React (TypeScript), using Vite for bundling and development. It is designed to be lightweight and accessible via standard web browsers, without requiring installation. As a proof-of-concept prototype, its core functionalities include persona generation and selection, persona-specific exploration of privacy harm scenario, and design diagnosis embedded in lo-fi user flows. All functionality is implemented on the frontend, and the app makes stateless API calls to LLMs for narrative generation.

To support consistency in user studies, the system currently relies on a fixed set of pre-generated, hard-coded user flows that serve as inputs for design analysis. While this limits immediate generalizability, the system was developed to explore core interaction patterns and content framing strategies under controlled conditions. Future versions are intended to support real-time harm analysis on dynamically uploaded user flows. This would require additional engineering effort but is feasible within the current architecture.

\vspace{2em}

\section{User Study: Evaluating PrivacyMotiv's Impacts through Privacy UX Review}

To evaluate the effectiveness of \textit{PrivacyMotiv} in enhancing designers' intrinsic motivation for reflective privacy-aware design through empathy building, we conducted a counterbalanced, within-subjects study with professional UX practitioners. Our primary goal was to investigate the impact of LLM-generated persona journeys on designers' motivation to proactively engage in privacy-aware design, as well as their empathy toward users affected by privacy incidents. This investigation is structured around the following four evaluation questions (EQs):

\begin{itemize}
    \item \textbf{EQ1 (Empathy):} How does \textit{PrivacyMotiv} influence designers' empathy toward users with vulnerabilities to privacy incidents? 
    
    \item \textbf{EQ2 (Motivation):} What is the impact of \textit{PrivacyMotiv} on designers' intrinsic motivation to proactively engage in privacy-aware design?
    
    \item \textbf{EQ3 (Review Outcomes):} Compared to their self-proposed review methods, how does \textit{PrivacyMotiv} affect designers' review results of identified privacy-invasive design problems and alternative design suggestions?
    
    \item \textbf{EQ4 (Usage Patterns):} How do designers use and perceive about \textit{PrivacyMotiv}, and what are their intentions to adopt it in real-life professional practices?
\end{itemize}

\subsection{Participants}

We interviewed 16 professional UX practitioners (P1--P16; see Appendix~\ref{appendix:participant-details}). This sample size aligns with prior within-subjects HCI studies focusing on qualitative insights and behavioral
changes~\cite{10.1145/3613904.3642794, li_honeysuckle_2021}. As summarized in~\autoref{tab:participant-summary}, participants ($\text{N}=16$) represented a diverse mix of UX professions (e.g., designers, researchers, strategists), spanning the broader UX discipline and working primarily in mid-to-large companies ($\text{62.5\%}$ in organizations with 5k+ employees). On average, they reported 4.5 years of total professional experience (estimated using the midpoints of self-reported ranges), reflecting practitioners commonly responsible for routine, day-to-day design delivery.

All participants were familiar with user flows, underscoring their role as a core coordination and review tool in UX practice and supporting our decision to use them as system input. Recruitment involved outreach through an author’s professional network and a professional UX design community's Slack channel, targeting experienced practitioners while ensuring relevance to real-world design work. We prioritized diversity in experience level, company size, and design specialization. Importantly, participants were not informed of the study’s specific aims or hypotheses, minimizing bias in their responses. Participation inclusion and exclusion criteria are detailed in Appendix~\ref{appendix:participation-criteria}.


\begin{table}[hbtp]
\centering
\small
\setlength{\tabcolsep}{4pt}
\begin{tabular}{@{}p{0.45\columnwidth} p{0.55\columnwidth}@{}}
\toprule
\textbf{Category} & \textbf{Distribution / Value} \\ 
\midrule
Number of Participants & 16 \\ 
\midrule
UX Profession & 9 Product/UX Designers \\
              & 3 UX Researchers \\
              & 2 UX Strategists \\
              & 1 Visual \& Interactive Designer \\
              & 1 Technical Design Engineer \\
\midrule
Avg. Total Experience    & 4.5 years (based on range midpoints) \\
Total Exp. Ranges (n=16) & 1--3 yrs (1), 3--5 yrs (13), 6--9 yrs (2) \\
\midrule
Company Size Distribution & 5k+ (62.5\%) \\
                          & 1k--5k (25.0\%) \\
                          & 201--1k (12.5\%) \\
                          & 1--200 (0\%) \\
\bottomrule
\end{tabular}
\caption{Summary of participant roles, professional experience, and company size. Values are based on pre-session survey responses from all 16 participants (see Appendix~\ref{appendix:participant-details}).}
\label{tab:participant-summary}
\end{table}



\subsection{Study Design}
We evaluate our LLM-powered tool, \textit{PrivacyMotiv}, against designers' self-proposed methods following a within-subjects study design.
A within-subjects study design controls for individual differences in designers' baseline empathy, privacy awareness, and professional experience, which could otherwise confound the results~\cite{li_honeysuckle_2021, wang2022documentation}.

Each participant was asked to complete two privacy review tasks: one for each condition on a different application scenario. We define a \textbf{privacy review} as a similar activity of UX design review but focus on identifying potentially privacy-invasive design problems and proposing alternative, privacy-preserving suggestions. 

To mitigate order and learning effects, we counterbalanced the order of both the conditions and the application scenarios across four participant groups.
Each participant was randomly assigned to one of them.

\begin{itemize}
    \item \textbf{G1:} \textit{Self-Proposed Approach} (AppA) $\rightarrow$ \textit{PrivacyMotiv} (AppB)
    \item \textbf{G2:} \textit{PrivacyMotiv} (AppA) $\rightarrow$ \textit{Self-Proposed Approach} (AppB)
    \item \textbf{G3:} \textit{Self-Proposed Approach} (AppB) $\rightarrow$ \textit{PrivacyMotiv} (AppA)
    \item \textbf{G4:} \textit{PrivacyMotiv} (AppB) $\rightarrow$ \textit{Self-Proposed Approach} (AppA)
\end{itemize}

\subsubsection{Experimental Conditions}

\paragraph{\textbf{Baseline: Self-Proposed Approach}}
As structured privacy reviews are not yet standard practice in UX design (supported by~\autoref{sec:context_results}), we established a realistic baseline condition. Participants were asked to conduct the privacy review using any approach that felt natural to their professional experience. This included applying design principles, considering user needs, or drawing from general work experience (specific approaches:~\autoref{subsec:usage_pattern}). This flexible approach allows us to capture the diversity of privacy-related reasoning currently found in industry practice.

\paragraph{\textbf{Experimental: PrivacyMotiv}}
In the experimental condition, participants used \textit{PrivacyMotiv} to conduct their review. The tool provided materials designed to foster perspective-taking and empathy, including speculative user personas, journey narratives illustrating potential feature misuse, and analyses of potential privacy harms. These materials were generated with an LLM pipeline (see Section~\ref{subsec:story} for generation details) and were intentionally not manually edited. This allowed us to test the LLM's generation capabilities within the study's context.

\subsubsection{Fictional Applications and Materials}
\label{subsubsec:fictional_apps_and_materials}

We developed two fictional applications for the privacy review tasks, a practice consistent with prior research~\cite{li_honeysuckle_2021}. This approach allowed us to control complexity, test similar types of privacy risks across multiple application contexts, and ground the tasks in concrete, relatable examples of how social transparency features can raise unexpected privacy issues. Furthermore, using fictional apps helps mitigate potential biases stemming not only from participants’ prior attitudes and experiences with real applications, but also from the LLM’s pre-existing knowledge of them. Each application was presented as a set of early-stage design materials, including a design brief and lo-fi wireframes in Figma (see Figure ~\ref{fig:figma_overview} and ~\ref{fig:wemusic_flow_example} in Appendix~\ref{appendix:app_details}),  simulating the externally prepared artifacts typically reviewed in early-stage design practice.

The fictional applications were designed according to four criteria. First, for ecological validity, they were grounded in popular real-world applications to reflect genuine user needs. Second, for privacy relevance, we selected features with a documented history of unintended privacy harms, each verified against at least ten distinct real-world incidents reported across news media, social forums (e.g., Reddit), and app store reviews. Third, to mirror industry practice, the task scope was limited to a single feature comprising four functions. Finally, the apps represented different domains to reduce carryover effects. Based on these, we chose two real-world references, Spotify ``Friend Activity'' feature and Citizen ``OnAir'' livestream feature, to inspire the design of two fictional apps for our study: \textbf{App (A): WeMusic ``Friend Activity''}, and \textbf{App (B): NeighborNet ``LIVE+''}. Detailed descriptions of each app and feature see Appendix~\ref{appendix:app_details}.

\subsubsection{Procedure Overview}

Each study session lasted approximately 90 minutes and was conducted remotely via Zoom. Participants used their own devices and interacted with materials via shared Figma files and a web-based version of \textit{PrivacyMotiv}, simulating a realistic remote design setting.
We adopted a within-subjects design where each participant completed two privacy review tasks under different conditions (\textit{Self-Proposed} and \textit{PrivacyMotiv-assisted}), with task order counterbalanced. 
In each condition, participants were tasked to review one fictional app feature and identified at least one privacy issue and one corresponding design suggestion within a 25-minute timebox, a duration informed by pilot testing and participants' own reports of typical design review cadences~\autoref{sec:context_results}. They completed a short evaluation survey at the end of each task.
Participants followed a think-aloud protocol throughout, and sessions concluded with a semi-structured interview comparing the two conditions.

This study protocol was approved by the Institutional Review Board of Northeastern University (IRB\#25-05-32). Participants received \$75 as compensation for the 90-minute session, a rate consistent with compensation practices in prior CHI studies involving professional designers~\cite{palani2022don, solyst2025conduit}. A full description of the study procedure is provided in Appendix~\ref{appendix:user_study_procedure}.

\subsection{Data Collection and Analysis}
\label{subsec:data_analysis}

We used a mixed-methods approach, collecting quantitative data from validated psychometric scales  and qualitative data from participants' written review outcomes and think-aloud protocols, with qualitative findings used to interpret and contextualize patterns in the quantitative measures.

\paragraph{\textbf{Inter-Task Questionnaires (EQ1, EQ2, EQ4).}}
After each of the two review tasks, participants completed a questionnaire adapted from established psychological scales (Appendix~\ref{appendix:inter_task_survey}). All questions using a 7-point Likert scale (1 = Strongly Disagree, 4 = Neutral, 7 = Strongly Agree). To ensure the measures were ecologically valid for the specific domain of privacy UX, we followed established methodological practices in UX research, adapting the wording of original items to refer explicitly to the privacy review context (e.g., changing this activity'' to this privacy review process'') while preserving the semantic meaning of the core constructs~\cite{perrig2024measurement, tyack2020self}.
We assessed scale reliability using Cronbach’s alpha~\cite{tavakol2011making}. All three scales showed good internal consistency.



To measure \textit{empathy (EQ1)}, we utilized a 6-item scale synthesizing dimensions from two frameworks. We selected items from the Interpersonal Reactivity Index (IRI)~\cite{davis1983measuring} to assess the participant's capacity to adopt the user's perspective (cognitive empathy) and feel concern for potential user harms (affective empathy). As design empathy specifically aims to support ``ideation and decision making''~\cite{mattelmaki2002empathy}, we supplemented the IRI with Empathy Probe items~\cite{10.1145/2702123.2702466,mattelmaki2002empathy} to measure how empathic understanding translates into a desire for design action (Cronbach’s $\alpha = .827$).

For \textit{intrinsic motivation (EQ2)}, we adapted a 7-item scale based on the Intrinsic Motivation Inventory (IMI)~\cite{ryan1983relation}, consistent with its widespread use in HCI evaluation~\cite{tyack2020self}. We selected items that targeted the participant's engagement, perceived competence in performing the analysis, and the perceived value of the review activity (Cronbach’s $\alpha = .908$). 

To measure \textit{perceived usefulness (EQ4)}, we developed a 5-item task-level scale. Drawing on the System Usability Scale (SUS)~\cite{brooke1996sus} and the Technology Acceptance Model (TAM)~\cite{davis1989tam}, we created a shorter measure tailored to the privacy review context. The scale was designed to capture participants’ immediate judgments of each approach’s perceived utility and their intention to adopt similar practices in professional design work (Cronbach’s $\alpha = .941$).

\paragraph{\textbf{Review Outcomes (EQ3).}}
\hypertarget{para:review-outcomes}{}
To analyze the written review outcomes, we developed a rigorous qualitative coding scheme. For \textit{specificity}, we adapted Garrett’s five planes~\cite{garrett2022elements}, a model classifying user experience across five levels of abstraction: \textit{Strategy} (L1, high-level concept or feature decision), \textit{Scope} (L2, functions or options that support features), \textit{Structure} (L3, user flows and timing), \textit{Skeleton} (L4, page layout and placement), and \textit{Visual} (L5, specific UI components and text). We applied a ``highest-level'' decision rule, coding each item at its most concrete plane. For example, “Place a red color highlighted toggle on the Share screen” is coded as L5, not L4. For \textit{thematic content}, we adapted the seven principles of Privacy by Design (PbD)~\cite{cavoukian2009privacy}: Proactive not Reactive; Privacy as the Default; Privacy Embedded into Design; Full Functionality; End-to-End Security; Visibility \& Transparency; and Respect for User Privacy. Following a ``best-fit'' assignment rule, each issue was coded with the single principle that best captured its core. The complete codebooks are in the appendix (Tables~\ref{tab:spec-levels} and~\ref{tab:theme-defs}).

To ensure reliability, three researchers first collaboratively developed the codebook by iteratively discussing 25\% of the data to refine the definitions. Once the codebook was finalized, two authors independently coded the entire dataset. We calculated inter-rater reliability separately for each scheme, finding strong agreement across both. For the \textit{specificity} levels, the independent coding yielded a Cohen’s Kappa of $\kappa=0.86$. For the \textit{thematic content} (PbD principles), the process resulted in similarly high reliability ($\kappa=0.82$). Any remaining disagreements were resolved through discussion to ensure the final accuracy of the results. We analyzed the final coded data quantitatively by comparing the frequencies and distribution of codes between the two conditions.

\paragraph{\textbf{Think-Aloud Recording Analysis (EQ1--4).}}
Finally, to supplement all findings, two researchers conducted a qualitative thematic analysis of the transcripts of the think-aloud recordings.
Findings were discussed iteratively with the entire team.
No inter-rater reliability was calculated because the goal was to identify new themes rather than seeking agreement and validating a taxonomy.
This analysis provided in-depth context and explanatory insights into participants' survey responses and behaviors across all four evaluation questions.

\section{Results}


\subsection{Context Results: Design Review and Privacy Review in UX}
\label{sec:context_results}
To contextualize the impact of \textit{PrivacyMotiv}, we first characterize participants’ existing design review practices and how privacy concerns currently surface within them. We find that design reviews are frequent and valued, but their routines lack triggers that reliably activate privacy-oriented reasoning. Privacy work tends to be reactive and delegated rather than preventative and designer-led .

\paragraph{\textbf{Design Review as a Frequent but Informal Practice}} 
Design review was a routine practice for nearly all participants. 13 out of 16 participants reported weekly or bi-weekly cadences, often with multiple touchpoints per week. While terminology varied across companies (e.g., ``design review,'' ``crit/critique,'' ``playback''), the core purpose converged on rapid, multi-perspective feedback: \textit{``sometimes there are two or three design reviews in a week---it’s how we move work forward''} (P10). 
Sessions typically spanned the full project lifecycle (from discovery to pre-launch), involving UX designers, researchers, and PMs by default, with developers or marketers joining as needed. Many participants (12/16) described reviews as casual screen-share walkthroughs with live Q\&A or Figma comments (e.g., \textit{``peers just come and offer different perspectives''}, P4). 4 participants cited structured methods (e.g., Nielsen's heuristic checklists, IBM's ``video{+}feedback'' playbacks), which generally occurred at higher-fidelity stages. 
Sessions were typically time-boxed and fast-paced, with a median of 10–30 minutes per project or per person (e.g., \textit{``ten-minute silent reviews''}, P4; \textit{``fifteen minutes for my part''}, P10). Feedback quality is often attributed to reviewer seniority (e.g., P5 finds PMs triaging quickly and provide more valuable feedbacks than peer designers), while early-mid career designers needed more scaffolding. These findings underscore that design review is a highly valued, fast-paced, and informal practice, yet also highlight a gap for lightweight, structured review methods, especially during earlier design phases.

\paragraph{\textbf{Design Review Rarely Centers Privacy}} Internal designer review sessions concentrate on visual consistency and design-system adherence, interaction clarity, accessibility, usability, and even microcopy tone, like P9 states \textit{``we always start from the [UI component] system—spacing, tokens, components—before anything else''}. Cross-functional reviews foreground feasibility and sequencing (sprint fit, tech-debt), performance/reliability, analytics impact (conversion, retention), and launch risk (e.g., \textit{``engineering asks, `can this ship in the next sprint, and what’s the risk?''}, P1; \textit{``in playback we talk numbers—why retention or conversion is low''}, P7). Leadership/external touchpoints probe strategic fit, value proposition, and brand alignment.

\paragraph{\textbf{Privacy in Practice: Reactive and Delegated}} When asked directly whether they had ever conducted a privacy review, all sixteen participants answered \emph{no}. However, 11 of 16 participants described encountering privacy-related gates or concerns in their broader workflows, most often in cross-functional forums (e.g., stakeholder/engineering playbacks) rather than during standard design reviews. 
Across participants, privacy concerns were rarely addressed proactively or led by designers. For example, P2 noted that while their team had a privacy-related checklist, it was neither specific nor comprehensive; instead, privacy was treated as a secondary concern within broader design requirements, often overshadowed by accessibility. Similarly, P9 recounted discovering privacy issues only \emph{after} harms had occurred, surfacing through user complaints routed to the marketing team. It was only after these were raised in an organization-wide meeting that the design team was prompted to retrofit the design. 
Another common pattern was reliance on PMs or senior colleagues. As P10 described: \textit{``my project manager can see a lot [potential privacy-invasive issues] \ldots{} because they have connections with the legal team.''} Overall, participants engaged with privacy only superficially---typically reactively, and often through others' lines of responsibility. A few participants from industry-scale organizations (e.g., P11, P7) mentioned the presence of dedicated legal or ethics departments, but noted that these stakeholders remained peripheral to routine design review. Occasional privacy or security workshops were optional and focused on general awareness rather than project-specific UX decisions.

\subsection{Usage Pattern: Self-Proposed Approaches \& PrivacyMotiv (EQ4)}
\label{subsec:usage_pattern}
We compared how participants approached the privacy review task under the baseline (self-proposed) condition versus the \textit{PrivacyMotiv} condition, focusing on dominant strategies participants used in each condition and the practical limitations they articulated while performing the task.

\subsubsection{\textbf{Self-Proposed Approaches: Ad-hoc and Context-Poor}}
\label{subsubsec:self_propose_approaches}
In the baseline condition, participants used their preferred review methods, which we categorized into four types:

\begin{itemize}
    \item Bystander's Perspective (6 participants): the most common approach involved participants acting as detached professional evaluators who scanned design files using UX heuristics and design principles to identify flaws in control and interaction patterns.
    \item ``Imagine Myself as the User'' (5 participants): participants simulated users' perspective and identified potential privacy discomforts based on personal empathy and experiences, as P7 put it: \textit{``gut check what it feels like.''}
    \item External AI Assistant (4 participants): these participants used conversational AI tools like ChatGPT and Claude.ai. P1 provided ChatGPT-4o with screenshots of all four user flows and a detailed prompt. P14 used ChatGPT's voice function to describe the app and request general guidance. P6 and P12 used Claude.ai, sharing specific screenshots of problematic interfaces or flows to get targeted feedback.
    \item Collaborative Walkthrough (1 participant): one participant engaged in real-time discussion with the study moderator to uncover issues.
\end{itemize}

Given that design reviews are routine but rarely center privacy (\autoref{sec:context_results}), participants’ self-proposed baseline approaches reveal how real-life designers would navigate privacy reviews without structured support. A recurring limitation was the lack of context to anticipate harms beyond personal experience, particularly for vulnerable or non-typical users. P12 noted that without specific personas, \textit{``I might not have considered some users.''} Several participants described feeling unprepared and lacking structure; 
P16 compared it to \textit{``solving a math problem, but without a good calculator.''} Those using external AI also described friction in providing adequate context. P1 remarked, \textit{``I need to take a screenshot... then copy paste into my prompt... it's a lot of work.''} Overall, baseline approaches were characterized by ad-hoc reasoning, limited contextual grounding, and substantial effort to externalize relevant details.

\subsubsection{\textbf{PrivacyMotiv Usage: Structured and Multi-strategy}}
\label{subsubsec:privacymotiv-usage}
Compared to self-proposed approaches, participants engaged with \textit{PrivacyMotiv} in more structured patterns, developing strategies to surface the most salient information. We observed three primary patterns:

\begin{itemize}
    \item Narrative-First: most participants (11 of 16) prioritized understanding the user context---spending time on Persona Info, Story Description, and Harm \& Consequence sections to build empathy before examining the UI. As P7 noted, the story ``\textit{is very helpful for a designer... to feel the users}''. 
    \item UI/Flow-First: other participants (5 of 16), particularly those with stronger design focus, began with Persona Usage Flow and Design Diagnostics to unpack failure mechanisms. As P4 explained, ``\textit{[Privacy] harms \& consequences come from the flows}''.
    \item Triage Approach: many participants (12 of 16) from both groups also adopted an efficient triage strategy, using Persona Type labels (e.g., ``Domestic Violence Survivor'') to jump directly to the Harm \& Consequence section to assess severity quickly.
\end{itemize}

These strategies indicate that \textit{PrivacyMotiv} supported both depth-oriented and efficiency-oriented engagement: participants either built contextual understanding before evaluating UI or used flow\slash diagnostic structures to rapidly identify failure points.

\subsection{How PrivacyMotiv Changes Designers’ Empathy and Motivation (EQ1, EQ2)}
\label{sec:empathy_motivation}

\begin{figure}[htbp]
    \centering
    \includegraphics[width=\columnwidth]{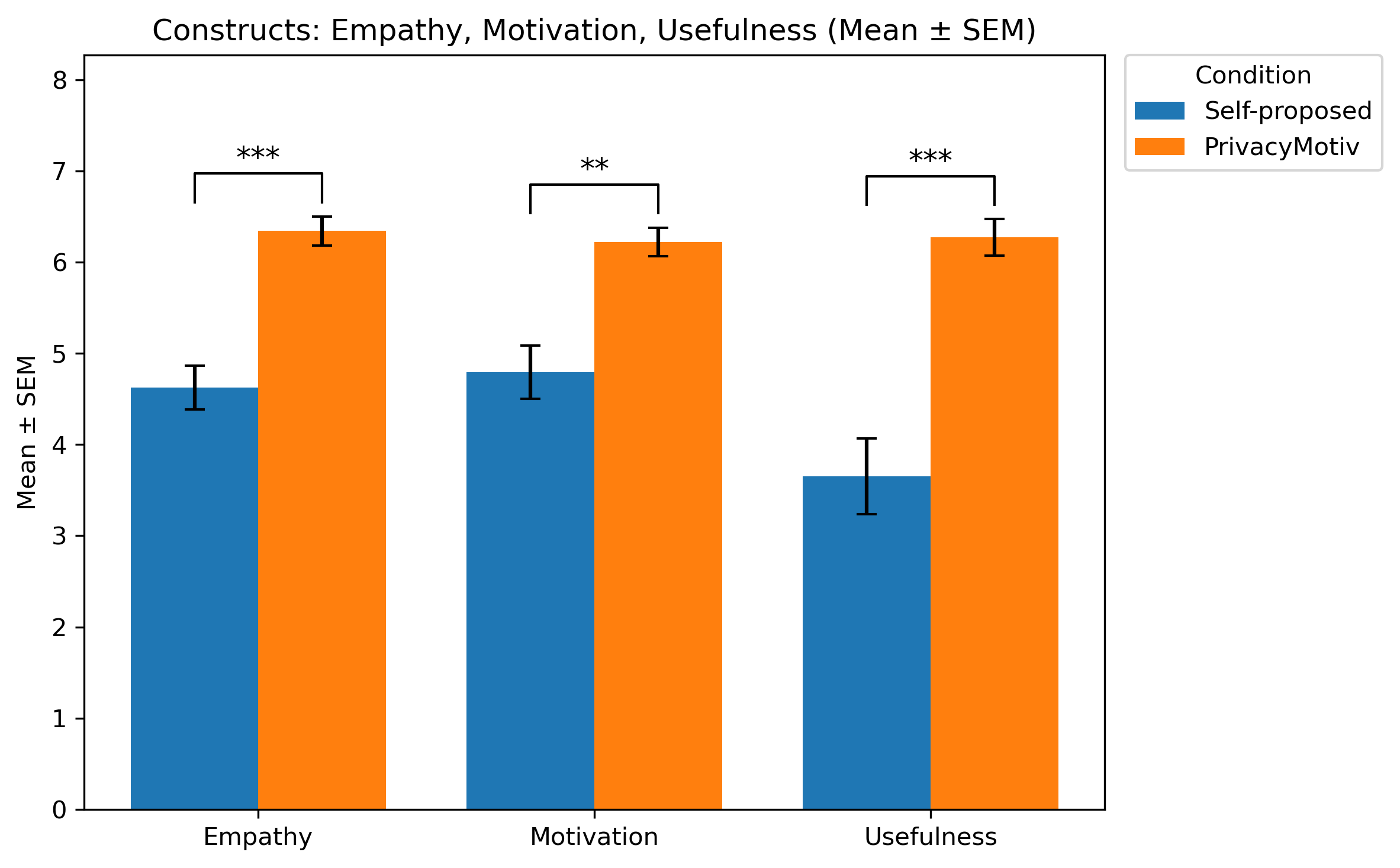}
    \caption{Comparisons of Empathy, Motivation, Usefulness between the designer’s self-proposed approach (baseline) and \textit{PrivacyMotiv}. Significant differences are indicated by asterisks based on paired t-tests with Benjamini-Hochberg correction (*$p$~<~0.05, **$p$~<~0.01, ***$p$~<~0.001).}
    \Description{A bar chart comparing mean scores of three constructs: Empathy, Motivation, and Usefulness between the Self-Proposed baseline condition and the PrivacyMotiv-assisted condition. Error bars indicate variability across participants. Asterisks above bar pairs denote statistically significant differences based on paired t-tests with Benjamini-Hochberg correction, with PrivacyMotiv scoring higher across all three constructs.}
    \label{fig:bar_constructs_primary}
\end{figure}

\subsubsection{\textbf{Quantitative Confirmation}}
We performed paired-samples $t$-tests to compare the general empathy, motivation, and usefulness between the self-proposed approach (baseline) and \textit{PrivacyMotiv} (\autoref{fig:bar_constructs_primary}). Three tests were performed in total and $p$-values have been corrected using the Benjamini-Hochberg (BH) procedure.

\begin{itemize}
    \item \textbf{Empathy (EQ1):} a paired-samples t-test showed that \textit{PrivacyMotiv} led to higher empathy scores ($M=6.3, SD=0.6$) than the baseline self-proposed approach ($M=4.6, SD=1.0$), $t(15)=5.4, p<.001, d=1.4$ (BH-corrected).
    \item \textbf{Intrinsic Motivation (EQ2):} a paired-samples t-test showed that \textit{PrivacyMotiv} led to higher motivation scores ($M=6.2, SD=0.6$) than the baseline self-proposed approach ($M=4.8, SD=1.2$), $t(15)=3.9, p=.002, d=1.0$ (BH-corrected).
    \item \textbf{Usefulness (EQ4):} a paired-samples t-test showed that \textit{PrivacyMotiv} led to higher perceived usability and usefulness scores ($M=6.3, SD=0.8$) than the baseline self-proposed approach ($M=3.7, SD=1.7$), $t(15)=5.1, p<.001, d=1.3$ (BH-corrected).
\end{itemize}

\subsubsection{\textbf{Qualitative Corroboration from Post-Task Reflections}}
Think-aloud protocols and post-task reflections corroborated these quantitative differences. Participants frequently described that \textit{PrivacyMotiv} made privacy concerns feel more salient and personally meaningful by grounding issues in concrete user contexts and consequences. For example, P15 stated that the tool heightened his ``privacy spidey sense,'' making him more attentive to privacy implications across interface actions. Participants also described the persona narratives as particularly effective for helping them connect interface decisions to user impact (e.g., \textit{``very easy for me to identify where the problem happened in the UI flow''} (P1)). We further synthesize these explanations into potential mechanisms in Section~\ref{sec:discussion}.

\subsection{How PrivacyMotiv Improves Review Outcomes and Actionability (EQ3)}
\label{subsec:privacy-review-outcomes}

\subsubsection{\textbf{Volume of Review Outcomes}}
\label{subsubsec:review-volume}
The \textit{PrivacyMotiv} condition generated a significantly higher volume of review outcomes. As detailed in Table~\ref{tab:summary_stats}, participants using \textit{PrivacyMotiv} identified \textbf{59\% more problems} and proposed \textbf{70\% more suggestions} on average than when using their self-proposed methods. This indicates that the tool’s scaffolding, particularly its use of persona-driven stories, was effective at prompting a more thorough review.


\begin{table}[h]
\centering
\small
\setlength{\tabcolsep}{4pt}
\caption{Review outcome statistics by condition.}
\label{tab:summary_stats}
\begin{tabular}{@{}p{0.48\columnwidth}cc@{}}
\toprule
\textbf{Metric} & \textbf{\textit{Self-Proposed}} & \textbf{\textit{PrivacyMotiv}} \\
\midrule
Total Problems Identified & 29 & 46 \\
Total Suggestions Proposed & 30 & 51 \\
\midrule
Avg. Problems per Participant & 1.9 & 2.9 \\
Avg. Suggestions per Participant & 2.0 & 3.2 \\
\bottomrule
\end{tabular}
\end{table}

\subsubsection{\textbf{Specificity of Outcomes}}
\label{subsubsec:specificity}
\begin{figure*}[t]
  \centering
  \includegraphics[width=0.88\textwidth]{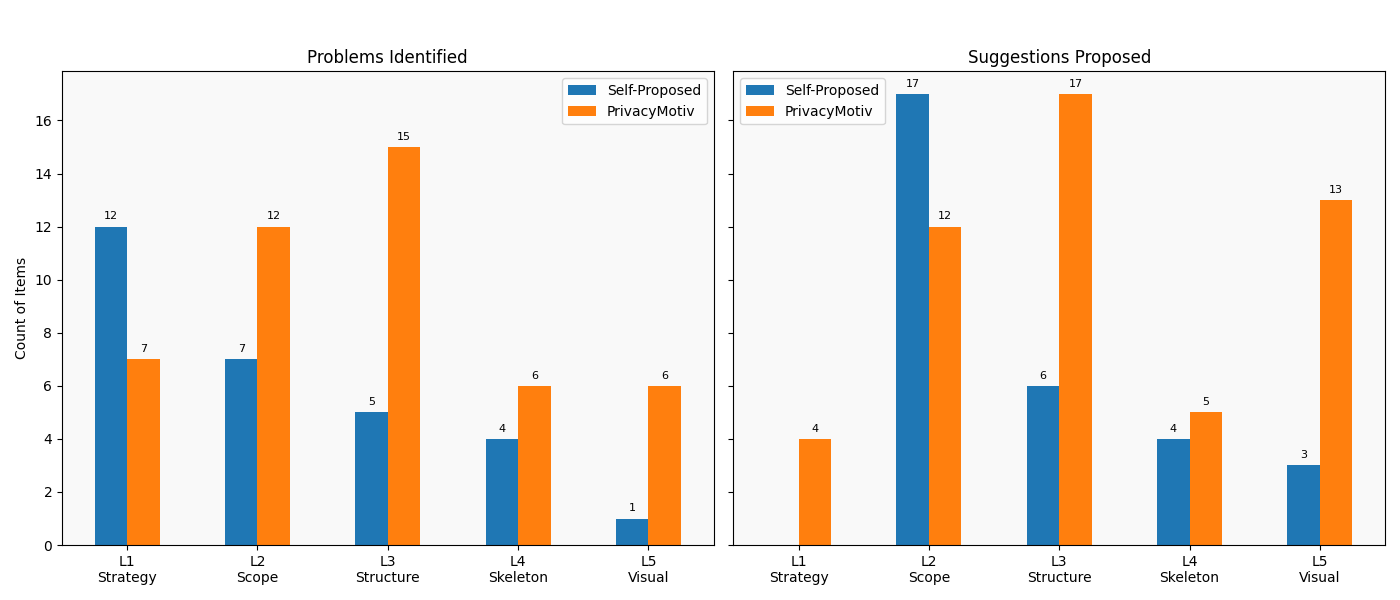}
  \caption{Specificity of problems and suggestions across Garrett's five planes (L1-L5). The Self-Proposed condition (blue) concentrated on abstract L1 problems and L2 suggestions. In contrast, PrivacyMotiv (orange) shifted focus toward more concrete layers, significantly increasing L3 and L5 outcomes while introducing strategic L1 suggestions absent in the baseline.}
  \Description{A grouped bar chart comparing the specificity of privacy problems and design suggestions across Garrett's five planes of user experience (L1 Strategy, L2 Scope, L3 Structure, L4 Skeleton, L5 Visual) between the Self-Proposed baseline condition shown in blue and the PrivacyMotiv-assisted condition shown in orange. The baseline condition concentrates on abstract L1 Strategy problems and L2 Scope suggestions, while PrivacyMotiv shifts distribution toward more concrete levels, notably increasing L3 Structure and L5 Visual outcomes and introducing L1 strategic suggestions that were absent in the baseline condition.}
  \label{fig:specificity}
\end{figure*}

Beyond sheer volume, the concreteness of outcomes differed substantially between conditions. Figure~\ref{fig:specificity} shows the distribution of problems and suggestions across~\hyperlink{para:review-outcomes}{\textit{Garrett's five planes}}~\cite{garrett2022elements}.

In the Self-Proposed condition, problems often remained at a high level of abstraction, focusing on \textbf{L1} (Strategy/Feature, 41\%) and \textbf{L2} (Scope/Function, 24\%). For example, P7 raised a concern about the high-level design decision of the LIVE feature, that it saves historical logs of live videos. They argued this strategic choice (L1) could inadvertently \textit{``encourage people to become influencers in their neighborhood,''} which might put the entire neighborhood at constant risk of surveillance. Suggestions in this condition were slightly more concrete, but still skewed towards high-level \textbf{L2} changes (Scope/Function, 57\%). For instance, P7's corresponding suggestion was to \textit{``include [a] reporting [function] to take down historic video from other's profile, if you are on camera in the video.''}  It remains a high-level concept rather than a practical fix to address the core concern about the feature's potential to normalize surveillance. Notably, no strategic (L1) suggestions were made, aligning with prior work suggesting designers may hesitate to challenge high-level decisions due to perceived lack of autonomy.

In the \textit{PrivacyMotiv} condition, we observed a clear shift toward greater concreteness across both problems and suggestions. Identified problems showed a clear move toward the concrete, transitioning from abstract strategies to concrete functions, and peaking at the \textbf{L3} (Structure/Flows, 33\%) level, where participants referred to specific flows, sequences, or timing issues. This contrasted sharply with the baseline, which identified almost no Visual (L5) problems (only 1 occurrence), whereas PrivacyMotiv successfully surfaced these concrete interface issues. For example, P10 highlighted confusion: \textit{``There is no warning before they start [a live broadcast], before they really understand what they are trying to do.''} Suggestions similarly trended more concrete than in the self-proposed condition, with notable increases at both \textbf{L3} (Structure/Flows, 33\%) and \textbf{L5} (Surface/Visual, 25\%) levels, indicating increased attention to specific interface components. Importantly, this shift does not suggest that participants focused solely on visual issues, but rather that they could translate abstract concerns into actionable design changes. Participants often articulated not only what functionality should change, but also where it should be applied, how it should be presented in the interface, and how users would perceive it. For instance, P6 offered a specific visual design critique (L5): \textit{``The `Private Session' button looks like a link to another page, [it] should be a toggle button, otherwise it could confuse users.''} Notably, while the Self-Proposed condition yielded more strategic-level (L1) problems, only \textit{PrivacyMotiv} prompted strategic suggestions. For example, after seeing a persona struggle with unwanted social connections, P7 proposed a strategic while actionable plan: \textit{``I wouldn't anticipate that [Facebook] syncing overrides this friend status in this app... I will push back [this design decision] on PM.''} This suggests that grounding practitioners in the concrete harms experienced by personas helped participants both drill down into concrete UI and flow-level details for actionable changes and reflect on foundational solutions.

\subsubsection{\textbf{Themes of Privacy Problems Indentified}}
\label{subsubsec:theme}
\begin{figure*}[t]
  \centering
  \includegraphics[width=0.85\textwidth]{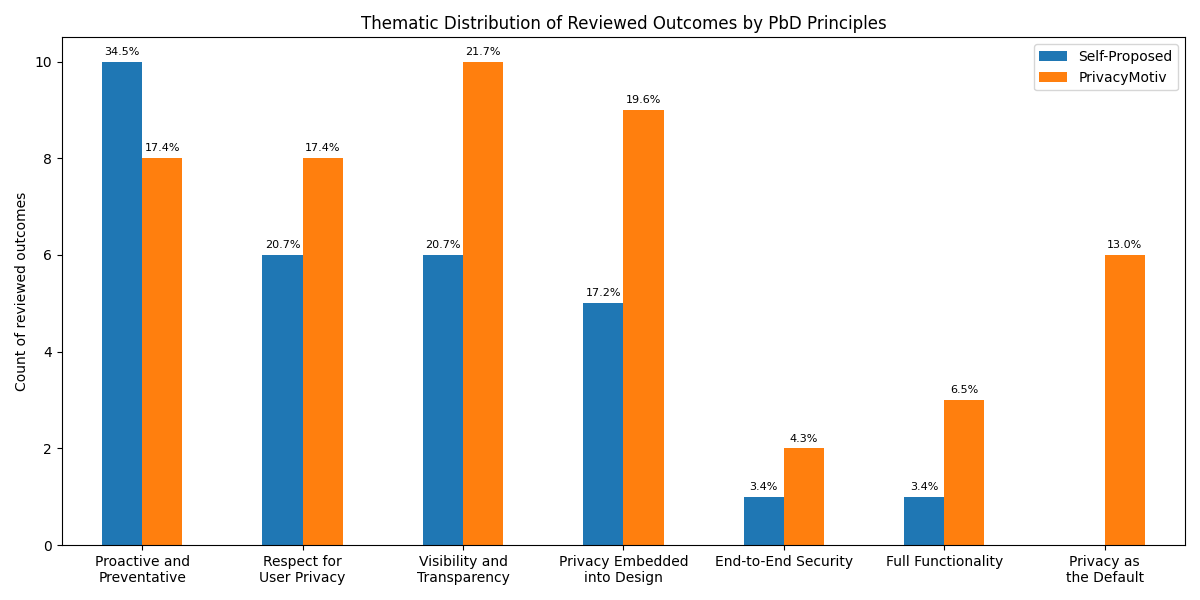}
  \caption{Thematic distribution of identified problems across Privacy by Design (PbD) principles. \textit{PrivacyMotiv} expanded coverage to include themes like Privacy as the Default.}
  \Description{A bar chart showing the thematic distribution of privacy problems identified by participants across Privacy by Design principles, comparing the baseline condition and the PrivacyMotiv-assisted condition. The chart shows that PrivacyMotiv expanded coverage across more PbD principles, notably surfacing Privacy as the Default as a theme that was absent in baseline reviews.}
  \label{fig:themes}
\end{figure*}

Our analysis of the outcome themes, structured by the~\hyperlink{para:review-outcomes}{\textit{PbD framework}}~\cite{cavoukian2009privacy}, reveals that PrivacyMotiv significantly broadened the scope of privacy issues practitioners considered (see Figure~\ref{fig:themes}).

First, \textit{PrivacyMotiv} drove a distinct shift toward Visibility and Transparency, which became the most frequently identified theme. This aligns with our specificity findings, suggesting \textit{PrivacyMotiv} enables designers to recognize privacy as a critical matter of user communication and interaction, rather than a mere functional data requirement.


Second, A substantial increase in problem identification occurred in Privacy Embedded into Design, suggesting that \textit{PrivacyMotiv} helped participants recognize deeply integrated issues previously overlooked. Although participants were sometimes unsure whether certain confusing UX elements qualified as privacy issues, \textit{PrivacyMotiv} prompted them to identify more elements with potential implications. For instance, P6 questioned a subtle UI element that might have gone unnoticed: (\textit{``On the UI, I didn’t see any function or button to remove someone… it just says ‘Follow’ or something. I guess if I press it, it becomes ‘Unfollow’?’’}). \textit{PrivacyMotiv} helped reveal hidden issues that intuition alone may miss.

Crucially, \textit{PrivacyMotiv} exposed violations of Privacy as Default, a category virtually absent in the self-proposed condition where standard UI patterns created a false sense of security. For instance, P7 dismissed a livestreaming flow as \textit{``very safe''} due to several consent pop-ups exist; PrivacyMotiv prompted the identification of missing upstream safeguards for vulnerable groups (e.g., minors), demanding protection before the broadcast began.

Finally, the decrease in Proactive not Reactive codes shows analytical maturation. Whereas control participants often labeled generic notifications as \textit{``preventative,''} (P1,4,6,7,10); \textit{PrivacyMotiv} encouraged identifying root causes. Consequently, surface-level prevention issues were reclassified as specific structural failures (e.g., Respect for User Privacy), demonstrating a more granular understanding of privacy architecture.

\subsubsection{\textbf{Originality of Suggestions Proposed}}
\label{subsubsec:originiality}
Beyond quantitative metrics, \textit{PrivacyMotiv} fostered confident suggesting, with participants frequently leveraging persona narratives to articulate why specific flaws constituted harms.  \textit{PrivacyMotiv} acted as a scaffold for originality. Participants consistently extended generic diagnostics into specific, novel features. For example, while \textit{PrivacyMotiv} offered a generic warning regarding WeMusic's stalking risks, P14 expanded this into a specific feature to \textit{``notify users of frequent profile viewers.''} Similarly, for NeighborNet, participants devised creative technical controls such as \textit{``automatically blur[ring] faces''} (P8) and \textit{``voice-changing''} for anonymity (P9), alongside procedural safeguards like parental supervision onboarding for minors (P13). Notably, these proposals rarely mirrored the tool's defaults. Instead, designers used \textit{PrivacyMotiv} as a springboard to bridge the gap between identifying abstract risks and proposing tangible, creative solutions.

\subsection{PrivacyMotiv Adoption Intention and User Mindset Shifts (EQ4)}
\label{subsec:readiness}

\subsubsection{\textbf{Perceived Strengths and Areas for Improvement}}
\label{subsubsec:strengths}
Participants evaluated \textit{PrivacyMotiv} for its fit within fast-paced design review routines, highlighting both enabling strengths and areas for refinement.

\paragraph{Strengths \& Advantages}
The tool's rich narrative context was praised as educational and empathy-building, heightening participants' ``privacy spidey sense'' (P15) and reframing technical errors as distinct ``harmful'' outcomes (P16). Participants also valued the tool's efficiency in automating scenario creation. Finally, the user flow with integrated diagnostics was celebrated as designer-friendly; the direct mapping of critiques to UI elements provided an intuitive structure comparable to resolving ``tickets'' in professional workflows (P11).



\paragraph{Areas for Improvement}
Participants also described several adoption barriers related to content realism, usability, and trust in underlying AI process. Some perceived personas as extreme edge cases, with P1 initially questioning the narratives \textit{``a little bit extreme''}. However, reading full narratives often convinced participants these cases feel more plausible and helped surface overlooked user needs in mainstream persona practices. Others critiqued high content density in the Persona and Story narratives, and some perceived redundancy across sections (e.g., Story vs.\ Usage Flow; Privacy Tensions vs.\ Harms). While some users mitigated this by ``triaging'', jumping directly from Usage Flows to Harms, they recommended streamlining for improved readability. A few participants expressed skepticism regarding AI. P3 cautioned that the tool's convenience might erode designers' critical thinking skills. However, our Review Outcomes (\autoref{subsec:privacy-review-outcomes}) suggest that critical engagement remained robust. Rather than simply following \textit{PrivacyMotiv}'s diagnoses, designers actively questioned the AI's output and synthesized original, creative solutions themselves. Finally, over half (9/16) desired persona customization. impressed by the tool's handling of vulnerable groups, they hoped to expand by inputting their company's own personas to better integrate privacy reviews into business-focused workflows.

\subsubsection{\textbf{Post-use Shifts in Privacy Attitudes and Understanding}}
\label{subsubsec:post-use shift}

Participants also described shifts in how they viewed privacy review responsibility and how they defined ``privacy'' in UX work after using \textit{PrivacyMotiv}.

\paragraph{Attitudes towards Privacy Review}
\label{sec:attitudes-towards-privacy-review}
Before using \textit{PrivacyMotiv}, a majority (11/16) endorsed privacy review in principle, yet many framed responsibility as diffuse or located elsewhere (e.g., PMs, legal, or security teams, as described by five participants), or as not central to UX design reviews (e.g., \textit{``[Privacy review is not the responsibility of designers because] I need to go through policies and requirements first''} (P11); \textit{``I think about ethical reviews more; I don’t think about privacy reviews as much''} (P13)). After using \textit{PrivacyMotiv}, most (13/16) expressed intent to integrate privacy considerations into routine reviews, preferring lightweight scaffolds that align with familiar artifacts (e.g., lo-fi wireframes) and typical timeboxes (10–30 minutes). As P15 put it, \textit{``I’ll start thinking more about these design choices around privacy.''} Others noted that structured prompts helped fill expertise gaps, especially for anticipating vulnerable contexts. 

\paragraph{Understanding of Privacy}
\label{sec:understanding-of-privacy}
Initially, participants often defined privacy in terms of compliance or attributes-permissions and specific data types (e.g., \textit{``it’s basically about what permission you’re giving to the app''} (P7); \textit{``privacy in my mind is \ldots{} users’ location''} (P1)). Several equated privacy with data sharing between entities (e.g., \textit{``previously, I think privacy is only about the data-sharing function \ldots{} sharing what data to which company''} (P16)). After using \textit{PrivacyMotiv}, participants’ descriptions broadened across three recurring themes: 12 of 16 articulated privacy in terms of concrete harms and lived consequences; 10 of 16 reasoned about privacy through contextual audiences and mechanisms---who sees what, when, and how (e.g., follower syncs, cross-surface exposure, resharing, persistence); and 9 of 16 emphasized vulnerable cases and bystander perspectives when explaining why an issue was privacy-related. Several explicitly contrasted pre- and post-use understandings; for example, P15 noted a move from focusing only on identifiable information to being \textit{``curious about every action taken on an interface,''} noting \textit{``There’s a lot of risk involved with the kinds of data being shared.''}

\subsubsection{\textbf{Further Implementation and Adoption Recommendations}}
\label{sec:adoption-recommendations}
Participants expressed optimism about adopting a tool like \textit{PrivacyMotiv} to augment, not replace professional judgment, while emphasizing the need to independently verify flagged issues (e.g., \textit{``verify information on my own''}, p4). Most favored using the tool early in the design process, after lo-fi wireframes but before high-fidelity mockups, to ``speed up this process.''

Practically, participants proposed embedding \textit{PrivacyMotiv} into existing workflows to ensure low-friction integration. Some suggested attaching reports to design critiques as shared reference points, while others envisioned using condensed prompts as lightweight checklists during sprint rituals. Several recommended linking the reports to tickets or PRDs to document privacy-related decisions and trade-offs. Direct integration into design platforms (e.g., Figma plugins) was also seen as a promising direction.

Beyond workflow integration, participants also highlighted its collaborative potential. Integrating it into conversations with legal teams was viewed as a way to streamline handoffs and clarify concerns (e.g., \textit{``make our collaboration with the company’s privacy experts more efficient''}, P11). Additionally, the narrative format was praised for helping communicate abstract harms to stakeholders such as PMs and clients, making potential risks more tangible and justifying prioritization.

\section{Discussion}
\label{sec:discussion}

Our findings reveal that designers struggle to identify unintended privacy harms, extending the literature beyond intentional manipulation~\cite{gray2018dark, deceptive_design, bosch_tales_2016}. We highlight a routine challenge: harms from well-intentioned intended UX designs often become visible only in specific social contexts. Below, we discuss: (1)~\autoref{subsec:discussion_1}, why current design reviews fail to activate privacy reasoning; (2)~\autoref{subsec:discussion_2}, how \textit{PrivacyMotiv} enables an empathy-to-action pathway (EQ1-3); and (3)~\autoref{subsec:discussion_3}, what our results imply for workflow integration, tool design, and research-to-practice translation (EQ4).

\subsection{Routine Design Review Misses Privacy: A Context Bottleneck, Not a Value Gap}
\label{subsec:discussion_1}
Design reviews are frequent and valued in industry practice, but our findings show these routines are optimized for speed, coherence, feasibility, and measurable outcomes (e.g., design-system adherence, usability, sprint fit, conversion metrics), not for anticipating harms (\autoref{sec:context_results}). Privacy concerns often appear as downstream exceptions, surfacing through user complaints, compliance gates, or escalation to PM/legal/security, rather than integrated into early critique. This echoes practitioner-facing accounts that privacy is difficult to routinize and unevenly distributed across roles \cite{zhang-kennedy_navigating_2024}, contributing to reactive rather than preventative privacy work. While prior usable privacy research has documented similar deferral patterns among software developers~\cite{6876252, 10.5555/3291228.3291251}, our findings extend this to UX designers specifically, whose upstream design decisions make early intervention particularly consequential.

Crucially, the baseline condition makes this gap concrete. Participants’ self-proposed strategies (\autoref{subsubsec:self_propose_approaches}) were not ``incorrect'', they were plausible but limited by a common bottleneck: privacy harms are inherently contextual---they depend on \emph{who} the user is, \emph{what} constraints they face, \emph{who} can observe them, and \emph{how} disclosures persist and propagate---and designers lack sufficient context to anticipate these harms beyond their own experience. General-purpose support (including off-the-shelf AI) still requires designers to author this missing context at high cost, which participants experienced as labor-intensive and uncertain. Taken together, these findings suggest a design opportunity: making context explicit and portable may help transform privacy from an occasional ``gate'' into a routine, critique-able dimension.

\subsection{Speculative Persona Journeys Enable Empathy-to-Action}
\label{subsec:discussion_2}
Lack of motivation is widely recognized as a barrier to adopting responsible design practices \cite{wang2024farsight, rakova2021responsible, madaio2020co}. Our results show significant improvements in empathy and intrinsic motivation when designers used \textit{PrivacyMotiv} (\autoref{subsec:data_analysis}), suggesting that speculative persona journeys can address this motivational gap while also improving the actionability of review outcomes (\autoref{subsec:privacy-review-outcomes}). We synthesize the qualitative data into three potential mechanisms explaining how this empathy-to-action pathway operates.

\paragraph{\textbf{Mechanism 1: Vulnerability-centered personas broaden the reviewer’s lens.}}
\label{subsec:discussion_2_m1}
In baseline reviews, participants often projected from ``typical user'' assumptions, limiting the range of anticipated harms. \textit{PrivacyMotiv} introduced vulnerability-centered personas that surfaced overlooked users and contexts, shifting attention from generic ``privacy settings'' toward \emph{who} is at risk and \emph{under what conditions}. For example, persona labels like ``Domestic Violence Survivor'' made visibility and audience boundaries immediately salient. Designers who anchored on such personas more frequently examined issues like default sharing, persistence, and discoverability, shifting their reasoning from attribute- or permission-centric to scenario-based privacy framings involving actors, audiences, persistence, and linkages (\autoref{sec:understanding-of-privacy}). Unlike generic personas, vulnerability-centered personas surface harms that are systematically invisible in mainstream design practice---precisely because they fall outside the ``typical user'' assumptions designers default to~\cite{sannon2022privacy, mcdonald2022privacy}.

At the same time, adoption feedback highlights a tension: some participants initially perceived personas as ``extreme edge cases'' (\autoref{subsec:readiness}), suggesting that such scaffolds must be carefully framed to avoid othering; one promising direction is to represent vulnerability as situational rather than static, aligning with socio-ecological framings that focus on contexts over stigmatizing categories~\cite{tang2025beyond}. This tension also reflects a deeper overlap between vulnerability and trauma: personas representing groups such as domestic violence survivors carry inherently distressing content that may induce vicarious distress in designers~\cite{10.1145/3491102.3517475}. The psychological effects of engaging with such content were not explicitly investigated in this study; future work should examine this through a trauma-informed design lens~\cite{10.1145/3491102.3517475}.

\paragraph{\textbf{Mechanism 2: Narrative sequencing encourages perspective-taking before UI judgment.}}
\label{subsec:discussion_2_m2}
Most participants followed a narrative-first strategy, reading persona context, story description, and harms before inspecting interface details. This sequencing (user $\rightarrow$ harm $\rightarrow$ UI) shifted what ``counted'' as relevant evidence: interface choices were interpreted through contextual constraints and consequences, not as generic usability issues.
The story prompted reasoning about how an interaction would unfold for \emph{this} user in \emph{this} context (e.g., what happens if a post persists; who can discover it; what traces remain), before ``judging'' the interface. 

Even when participants did not read the full narratives, many still used persona labels and harm summaries as triage shortcuts: some jumping directly to the Harm \& Consequence section to assess severity before tracing back to relevant touchpoints. As detailed in~\autoref{subsubsec:privacymotiv-usage}, this pattern illustrates how narrative framing supported both deep engagement and fast-paced review. Meanwhile, a smaller group (5 of 16), often those emphasizing structural analysis, began with the Usage Flow and Design Diagnostics to examine the mechanics of failure. Though less explicitly narrative-driven, this structure-first approach remained oriented toward user impact, often seeded by the persona label or prior narrative exposure. These varied strategies highlight \textit{PrivacyMotiv}’s flexibility in supporting multiple, perspective-driven entry points rather than enforcing a single reading path.


While fictional scenarios can meaningfully shift practitioners' values reasoning and design judgments in real-world practice~\cite{wong_using_2021, 10.1145/3491102.3517589, Chen2026}, they also carry some related risks: that emotionally vivid narratives may skew prioritization toward extreme scenarios over proportional risk judgment~\cite{Betsch_Haase_Renkewitz_Schmid_2015}, and LLM-generated content may further encode biases toward particular harm framings~\cite{10.1145/3544548.3580688}, leading designers to over-invest in non-representative scenarios. Our results suggest both were partially mitigated: participants who initially found narratives extreme often revised this judgment after engaging with the full story (\autoref{subsubsec:strengths}), and participants actively generated mitigations beyond tool prompts (\autoref{subsubsec:originiality}), indicating critical engagement with rather than uncritical acceptance of generated content.

\paragraph{\textbf{Mechanism 3: Harm framing converts privacy from abstract issues into moral stakes and professional responsibility.}}
\label{subsec:discussion_2_m3}
Participants described a shift from treating privacy concerns as abstract ``issues'' to interpreting it as concrete, material harms. This reframing often came with heightened vigilance (e.g., ``privacy spidey sense'') and an expanded sense of responsibility around defaults, visibility, and persistence. Importantly, the reframing was not purely affective: participants linked vivid harms to both ethical urgency \emph{and} design feasibility. Privacy became \emph{worth addressing} because it felt ethically salient, and \emph{possible to address} because harms were translated into actionable design elements. This connection helps explain why empathy and motivation aligned in our study: empathy without actionable support may remain passive, while checklists without context may feel mechanical or detached. By pairing narrative harms with flow- and UI-grounded diagnostics, \textit{PrivacyMotiv} supported both affective engagement and practical agency, motivating participants move from ``noticing harm'' to ``changing the interface.''

These mechanisms help explain outcome shifts observed in~\autoref{subsec:privacy-review-outcomes}. While in~\autoref{fig:specificity}, baseline reviews often raised high-level concerns (L1--L2), they rarely included concrete solutions, consistent with the sense that strategic decisions are ``already fixed'' and outside designers’ control. In contrast, \textit{PrivacyMotiv} enabled participants to decompose concerns into flow- and UI-level interventions (L3--L5), increasing operational specificity (e.g., changing defaults, timing, visibility and controls). The PbD analysis similarly shows a broadened lens: participants not only identified more issues related to Visibility/Transparency and Privacy Embedded into Design, but uniquely surfaced Privacy as the Default in~\autoref{fig:themes} which were absent from baseline reviews. These findings echo known gaps in practitioners’ narrow conceptions of privacy \cite{li2018coconut} and skewed application of privacy design principles \cite{tahaei2022understanding}, suggesting that situated narratives make overlooked dimensions like persistence or cross-surface propagation more visible and actionable.

Finally, despite structured diagnostics, participants proposed novel mitigations beyond tool prompts (e.g., face blurring, voice-changing, anti-stalking signals), indicating that structured support can reduce cognitive overhead while enabling creative, contextually grounded solutions.

\subsection{Implications for Embedding Proactive Privacy Review in UX Practice}
\label{subsec:discussion_3}
Our adoption findings suggest that designers are willing to integrate privacy into routine critique if it remains lightweight and aligns with existing cadence, artifacts, and collaboration patterns (\autoref{subsec:readiness}). This implies a pragmatic model: a repeatable add-on anchored to lo-fi artifacts and short timeboxes (10--30 minutes), with clear escalation paths when legal or policy issues arise. Framed this way, privacy critique becomes a form of design reflection, not a post-hoc compliance gate. To support this model, future tools should focus on four key areas:

First, minimizing the burden of authoring context is critical for adoption. Compared to general-purpose AI, participants found value in systems like \textit{PrivacyMotiv} that pre-bind persona information, flow steps, and UI evidence into a traceable artifact, whereas off-the-shelf AI tools were seen as burdensome due to the need to generate detailed prompts and screenshots. Every flagged harm should link to the implicated flows or UI elements, and each mitigation should reference the lever it modifies (e.g., default, visibility, timing). Furthermore, participants' use of persona narratives to justify specific critiques to stakeholders suggests that fictional persona journeys can function as lightweight evidence artifacts for organizational advocacy (\autoref{sec:adoption-recommendations}). More broadly, \textit{PrivacyMotiv} enacts a reframing of privacy from a compliance-oriented obligation to an interpersonal and empathic phenomenon, where what matters is who is affected, under what social conditions, and with what consequences~\cite{nissenbaum2004privacy}---a shift reflected in participants' own post-use understanding of privacy (\autoref{subsubsec:post-use shift}).

Second, domain-specific customization must be balanced with guardrails against representational harms. While participants expressed interest in using company-specific personas, such customization requires careful situational framing to avoid stereotyping~\cite{tang2025beyond}. Practical strategies include requiring contextual fields (e.g., persona label), balancing common and vulnerable profiles, and offering plausibility tuning for product relevance.

Third, a scan-to-dive interaction model is necessary to support both fast-paced critique and deeper reflection. Progressive disclosure, starting with concise summaries (e.g., persona label, top harms, risky steps) and expandable into full narrative and diagnostics, can maintain the empathy-to-action pathway while accommodating time constraints.

Finally, review outputs must facilitate organizational communication, not just individual reflection. Participants proposed embedding outputs into PRDs, using narrative artifacts to persuade stakeholders. Tools should generate portable justifications that align with cross-role workflows and decision-making norms.

Taken together, our findings suggest that designers are not disengaged from privacy. Instead, routine critique lacks the contextual scaffolds to surface privacy as a first-class review dimension. \textit{PrivacyMotiv} shows that speculative personas, paired with flow- and UI-grounded diagnostics, can translate abstract risks into situated harms and actionable design levers, making privacy review both feasible and motivating.

\section{Limitations and Future Work}
This study was conducted in a controlled setting, whereas real design reviews involve more complex organizational constraints, stakeholder dynamics, and longer iteration cycles. We intentionally used heterogeneous baseline strategies to reflect practice, but this limits attribution to specific comparison methods. Our outcome measures (counts, specificity, thematic coding) are proxies and do not confirm downstream implementation or user impact. Although the study setting was grounded in real-world practice (task durations, lo-fi app reproductions, and externally prepared artifacts), empathy and motivation were measured immediately post-task, capturing task-level reactions rather than durable practice change. Future work will explore these questions through longitudinal field deployments and consider applying \textit{PrivacyMotiv} to designers' own in-progress designs


Persona narratives also introduce representational risks. We intentionally defined persona types using a single salient attribute rather than combining multiple identity dimensions. This conservative choice helped avoid speculative ``mixed'' personas that are difficult to justify with direct literature support and could inadvertently encode stereotypes. However, this design also limits our ability to capture intersectional vulnerability---where race, class, gender, and other factors co-shape risks and lived experience in ways that are not reducible to any one axis~\cite{schlesinger2017intersectional}.
Future work will explore how to develop \textit{deliberately grounded} intersectional personas, informed by intersectional scholarship and validated through participatory or community-reviewed methods, to improve engagement without inviting dismissal and stereotyping. The overlap between vulnerability and trauma also raises an unmeasured limitation: engaging with distressing persona content may affect designers' psychological wellbeing, warranting future investigation through a trauma-informed design lens~\autoref{subsec:discussion_2_m1}.


Beyond representational risks, the faithfulness of LLM-generated narratives to real-life privacy experiences cannot be fully guaranteed, and generated scenarios may not capture how specific vulnerable users actually respond to privacy threats in practice. Our results suggest this risk was partially mitigated by participants' critical engagement---actively questioning outputs and proposing solutions beyond tool suggestions (\autoref{subsubsec:originiality}, \autoref{subsec:discussion_2_m2})---though community-based validation with members of represented groups remains an important direction for future work.

As noted in Section~\ref{subsec:system_scope_and_implementation}, current \textit{PrivacyMotiv} system relies on pre-generated user flows and personas to support study consistency. While this limits immediate generalizability, it enabled us to isolate framing and interaction effects. The system was also evaluated exclusively in multi-party social app contexts; how it generalizes to designs where privacy is less central, such as single-user tools, remains an open question that future work will address alongside real-time analysis, dynamic content uploads, and progressive narrative disclosure.



\section{Conclusion}

We present a novel structured privacy review approach leveraging LLMs to generate contextualized persona journeys, enhancing UX designers' empathy and motivation to identify unintended privacy harms. We evaluated this approach via \textit{PrivacyMotiv}, a proof-of-concept system, in a within-subjects study with 16 UX professionals. Compared to self-proposed approaches, \textit{PrivacyMotiv} significantly improved designers' empathy toward users with vulnerabilities, intrinsic motivation, and perceived usefulness. It also yielded more specific and thematically comprehensive review outcomes. These findings highlight the value of structured, empathy-driven support for integrating privacy into design practice, demonstrating a promising direction for motivating privacy-aware UX.

\begin{acks}
This work is supported in part by the National Science Foundation CNS-2426396, CNS-2426395, CNS-2618863, and CNS-2442221. Any opinions, findings, and conclusions or recommendations expressed in this material are those of the authors and do not necessarily reflect the views of the sponsor. We thank the anonymous reviewers for their thoughtful feedback, which helped strengthen this work. We would also like to thank all the participants for their time and valuable contributions to this study.
\end{acks}

\bibliographystyle{ACM-Reference-Format}
\bibliography{references_all}

\appendix 
\clearpage

\section{Example Generated Persona Journey Story}
\label{appendix:example-story}
We present an example to illustrate how process-oriented narratives surface privacy harms as consequential experiences, shaped by user vulnerability and design choices (see~\autoref{fig:Eva_overview}).

Eva is a 16-year-old girl and high-school student who is facing online bullying from classmates~\cite{balas2023cyberbullying, rizza2013social, obaidat2023investigating}. She has high tech comfort because she moderates Discord servers and edits videos for school clubs, making her confident with app settings but still vulnerable to social risks. This persona was constructed to reflect teens navigating digital spaces under conditions of peer surveillance and harassment.



Within \textit{PrivacyMotiv}, Eva’s experience is generated by combining three functions: \textit{Viewing Friend Activity}, \textit{Enjoying private listening via Private Session}, and \textit{Removing friends from the Friend Activity Feed}. The story describes how Eva opens the app's friend feed and notices that a classmate who has bullied her appears there. Wanting to listen without being watched, she activates the platform's ``Private Session'' mode to hide her listening activity, assuming this will keep her behavior hidden from peers. However, because the session expires silently after six hours (a design choice surfaced in step 3 of the flow \textit{``Enjoying private listening via Private Session''}), and the app resumes broadcasting her listening activity, including a playlist named after her school’s choir, it becomes visible again to followers without her realizing it. The next morning, she attempts to remove the classmate from her feed by unfollowing them, but due to a persistent Facebook-synced connection, her listening remains visible to them. That afternoon, the classmate teases her about the school-related playlist and her late-night listening, triggering ridicule and emotional distress. The harms she faces include: (1) reputational harm from peers mocking her late-night listening and school-related playlist, (2) psychological distress from the breach of what she believed was a private space, (3) autonomy loss resulting from the app overriding her privacy expectations, and (4) relationship strain due to reactivated bullying dynamics among classmates. Ultimately, Eva responds by withdrawing from the platform, losing access to the social connection and music discovery features she once relied on.

This example illustrates how \textit{PrivacyMotiv} uses structured narratives to enable designers to empathize with privacy harms as consequential experiences shaped by app behaviors and individual vulnerabilities (\textbf{\hyperlink{D2}{D2}}). The system's approach of mapping design-level decisions (e.g., silent expiry, persistent sync, poor feedback) to specific harm outcomes allows designers to empathize with harms that emerge not from malicious intent but from interaction design choices, through the lens of users with vulnerabilities (\textbf{\hyperlink{D3}{D3}}). These structured scenario narratives make privacy harms legible as lived consequences of design interactions over time.

\section{Fictional Applications Details} 
\label{appendix:app_details}

\paragraph{\textbf{App(A)}: WeMusic---Friend Activity} This fictional music app was inspired by Spotify's \textit{Friend Activity}, a feature exemplifies a "well-intentioned" design. Following a common social trend seen across major platforms like Apple Music and YouTube Music, it was developed in response to strong user demand (over 7,000 user votes) to enhance social connection and music discovery~\cite{spotifyCommunity, techcrunchSpotifyCommunity, spotifySupportFriendActivity}. Despite this popular and benign design, the feature inadvertently enables significant privacy issues. Real-world examples show it can facilitate unintended behaviors such as emotional surveillance through playlist monitoring or even location tracking by observing a user's listening patterns in their car~\cite{firm_alfalfa_6017_you_2024}. To capture this tension in our study, the \textit{Friend Activity} feature in \textit{WeMusic} was presented to participants with four specific functions: (1) view friends' activity feeds, (2) share playlists and add friends, (3) start private listening sessions, and (4) remove friends by unfollowing or blocking users.

\paragraph{\textbf{App(B)}: NeighborNet---LIVE+} This fictional civic engagement app was inspired by Citizen's \textit{OnAir Livestream}. The real-world feature demonstrates clear user value, having become a popular tool for millions of users seeking real-time community safety and local coordination. Despite this genuine utility, the design creates significant privacy risks. For example, participation patterns can lead to personal preference inference, while live broadcasts can cause real-time location exposure, inadvertently revealing sensitive details like users’ daily schedules, family situations, or health needs. To explore this tension, the \textit{LIVE+} feature in \textit{NeighborNet} was presented to participants with four specific functions: (1) discover nearby live streams, (2) comment and react in streams, (3) start a personal live stream, and (4) manage live stream history by hiding or deleting past videos.

\begin{figure*}[t] 
    \centering 
    \includegraphics[width=\textwidth]{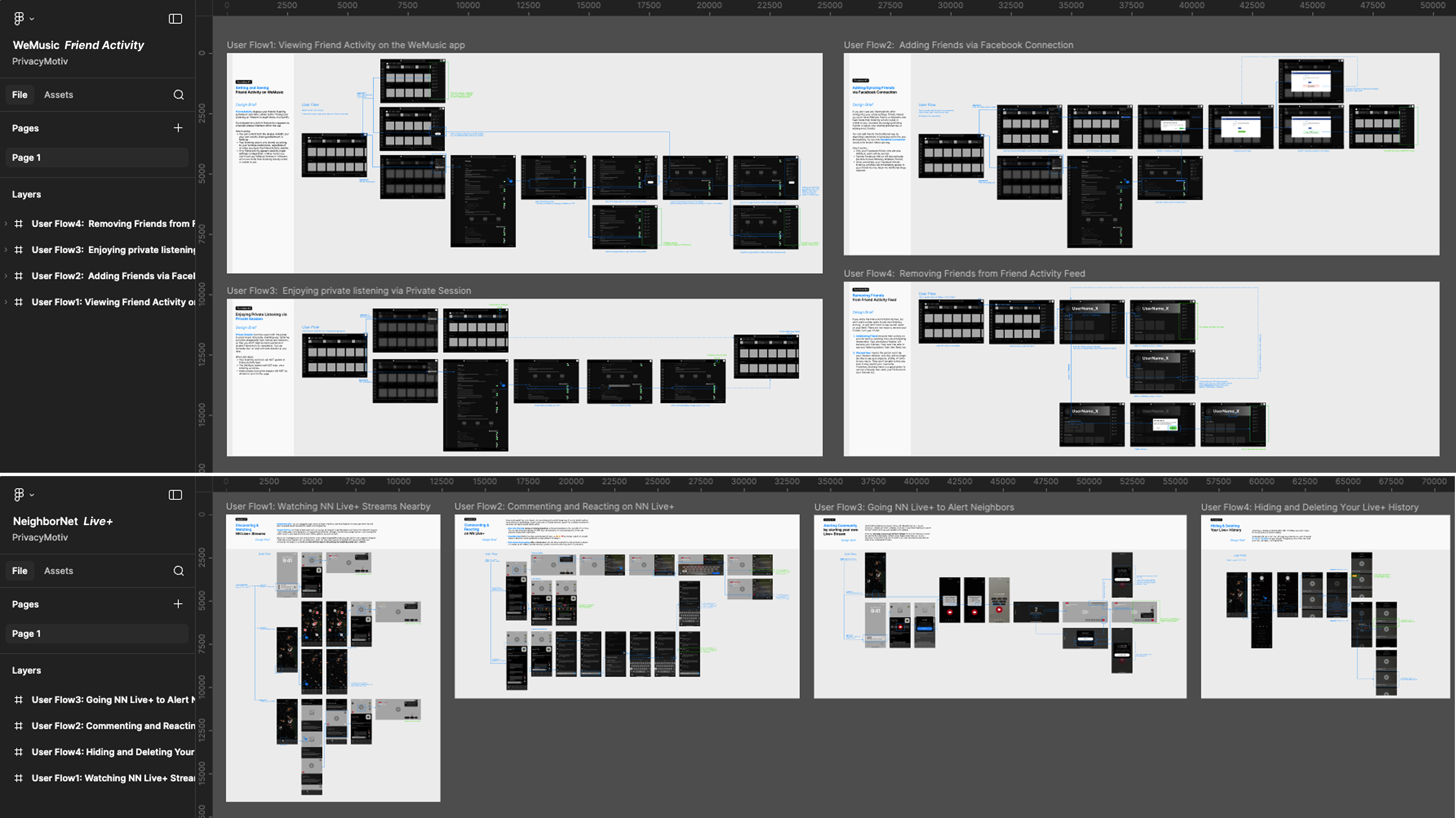}
    \caption{Overview of the Figma materials provided to participants for both App (A) and App (B).}
    \Description{An overview of the Figma workspace showing study materials for both applications, representing the set of materials participants reviewed during the privacy review tasks.}
    \label{fig:figma_overview} 
\end{figure*}

\begin{figure*}[t] 
    \centering 
    \includegraphics[width=1\linewidth]{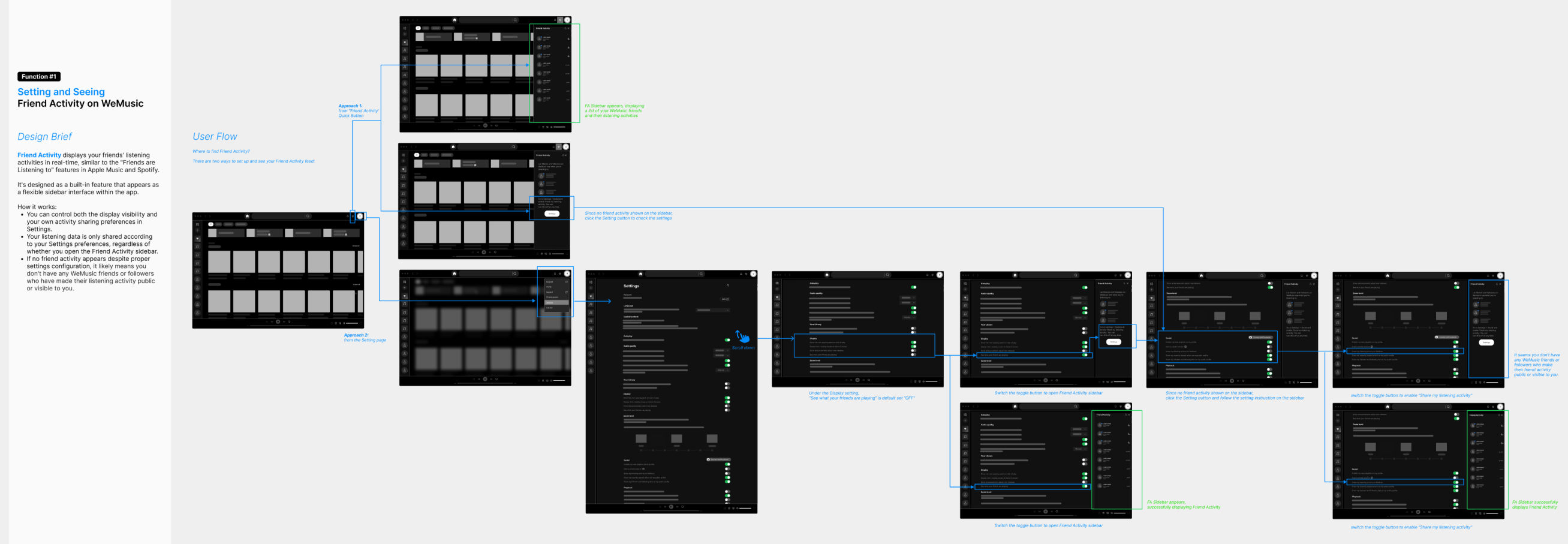} 
    \caption{An example of the WeMusic materials provided to participants. This board displays User Flow 1 (``Setting and Seeing Friend Activity''), comprising a design brief (left), lo-fi wireframes and flow arrows \& annotations (right) that illustrate how users configure system settings and access the Friend Activity sidebar.} 
    \Description{A Figma board example showing the WeMusic study materials for User Flow 1 (Setting and Seeing Friend Activity). The left side contains a design brief describing the feature context. The right side displays lo-fi wireframe screens connected by flow arrows, with annotations marking key interaction steps showing how users configure system settings and access the Friend Activity sidebar feature.}
    \label{fig:wemusic_flow_example} 
\end{figure*}

We developed a complete set of early-stage design materials in Figma for both fictional applications. Each Figma file set comprised four user flow boards, containing a short design brief, lo-fi wireframes, flow arrows and annotations, simulating a common design file used in a realistic UX workspace. Figure~\ref{fig:figma_overview} illustrates the full set of materials participants received for the review tasks. Additionally, Figure~\ref{fig:wemusic_flow_example} provides a detailed view of one single function flow from the WeMusic boards.

\section{Participation Details}
\subsection{Recruitment Criteria}
\label{appendix:participation-criteria}

\textbf{Inclusion criteria.} We required participants to have at least one year of professional UX design experience specifically within the U.S. market, as well as recent professional experience in UI/UX design within the past year (including current employment or freelance work), experience designing consumer-facing digital products or services, and familiarity with user research methods and design thinking processes. We focused on U.S.-based professional experience because cultural contexts and market conditions can significantly influence designers' motivations, empathy toward users, and approaches to user-centered design practices. 

\textbf{Exclusion criteria.} We excluded student designers without professional experience and individuals whose most recent professional UX design work occurred more than one year ago. We welcomed participants regardless of their prior exposure to privacy concepts or privacy-focused work experience, as we aimed to assess how PrivacyMotiv performs across the full spectrum of privacy awareness levels within the design community.

\subsection{Participants' details}
\label{appendix:participant-details}
\small

The seniority level for each participant is based on their official job title (e.g., "Senior UX Designer") as self-reported in the pre-session survey and defined by their respective employers, not by researcher assessment. This helps contextualize their experience relative to their company's internal structure. The full sample included 1 \textbf{Entry}, 11 \textbf{Junior}, and 4 \textbf{Senior} UX practitioners.

\begin{table}[t]
\centering
\scriptsize
\setlength{\tabcolsep}{3pt}
\begin{tabularx}{\columnwidth}{@{}l>{\raggedright\arraybackslash}Xllll@{}}
\toprule
Participant & Title & Company size & Total exp. & Seniority \\
\midrule
P1  & UX Designer & 201--1k & 3--5y & Jr. \\
P2  & UX Researcher & 5k+ & 3--5y & Jr. \\
P3  & Design Strategist & 5k+ & 3--5y & Jr. \\
P4  & Product Designer & 1k--5k & 3--5y & Jr. \\
P5  & UX Designer & 1k--5k & 6--9y & Jr. \\
P6  & Sr.\ UX Researcher & 5k+ & 3--5y & Sr. \\
P7  & Sr.\ Experience Designer & 5k+ & 6--9y & Sr. \\
P8  & UX Design Contractor & 201--1k & 1--3y & Entry \\
P9  & Product Designer & 5k+ & 3--5y & Jr. \\
P10 & Sr.\ UX Researcher & 5k+ & 3--5y & Sr. \\
P11 & User Experience Designer & 5k+ & 3--5y & Jr. \\
P12 & User Experience Designer & 5k+ & 3--5y & Jr. \\
P13 & Interactive and Visual Designer & 5k+ & 3--5y & Jr. \\
P14 & UX Designer & 1k--5k & 3--5y & Jr. \\
P15 & Senior Experience Strategist & 5k+ & 3--5y & Sr. \\
P16 & Technical UX Designer & 1k--5k & 3--5y & Jr. \\
\bottomrule
\end{tabularx}
\caption{Roster keyed to interview IDs (P1--P16). Experience fields are reported as survey ranges.}
\label{tab:one-roster}
\end{table}

\section{Study Procedure Details}

\subsection{Study Procedure}
\label{appendix:user_study_procedure}

All study sessions were conducted remotely via Zoom and lasted approximately 90 minutes. Participants used their own devices, simulating a realistic remote work environment. The fictional application materials were provided in shared Figma files, and \textit{PrivacyMotiv} was accessed through a web browser. To minimize fatigue, a short break was provided between the two main tasks. All sessions were audio and video recorded with participant consent.

Each session followed a four-step procedure for each participant:

\begin{enumerate}
    \item \textbf{Onboarding and Interview (20 minutes):} Prior to the study, participants were asked to complete a pre-session survey to collect basic demographic and professional information. The live session then began with a semi-structured interview to understand their professional backgrounds, familiarity with user flows, and experiences with design reviews in practice.

    \item \textbf{Privacy Review Tasks (25 minutes $\times$ 2):} Participants were sequentially assigned to one of the four counterbalanced groups. They performed two 25-minute privacy review tasks: one for each fictional application and under each condition.
    
    \paragraph{Scenario Setting.} For each task, we framed a realistic scenario by asking participants to imagine they were a UX designer working on a new, high-priority feature (Friend Activity or LIVE+) for an established product (WeMusic or NeighborNet). This framing mirrors a common industry practice where designers are tasked with designing features based on user research or business requirements, with the well-intentioned belief that they are serving user needs.
    
   \paragraph{Task Execution.} During the task, participants followed a think-aloud protocol to verbalize their reasoning and findings. They were required to document their findings, at least one privacy issue and one corresponding design suggestion, as free-form comments in a provided questionnaire. In the \textit{self-proposed approach} condition, participants could use any methods or external tools they chose and were asked to describe their process. In the \textit{PrivacyMotiv} condition, they were instructed to explore at least three different persona journeys to ensure sufficient engagement with the system.

    \item \textbf{Inter-Task Survey (5 minutes $\times$ 2):} After each of the two review tasks, participants completed a brief questionnaire measuring their empathy toward users, intrinsic motivation of privacy review, and the perceived usefulness of the approach they had just used.

    \item \textbf{Post-Study Debrief Interview (20 minutes):} The session concluded with a final semi-structured interview. We asked participants to compare their experiences and preferences between the two conditions, discuss any differences in how they considered users, their attitude and understanding of privacy, and share their general thoughts on incorporating privacy considerations into design practice.
\end{enumerate}

\section{Surveys and Interviews}

\subsection{Pre-session Survey}
\label{appendix:pre_session_survey}

\textbf{\textit{\\***Section A – Eligibility}}

\textbf{A1. Professional role (multi-select)}
\begin{itemize}
  \item Product/UX Designer
  \item UX Researcher
  \item Product Manager (with design responsibilities)
  \item Other (please specify): \underline{\hspace{2cm}}
\end{itemize}

\textbf{A2. Years of professional UX/UI design related experience (total)}
\begin{itemize}
  \item < 1 year (\textit{ineligible})
  \item 1–2 years
  \item 3–5 years
  \item 6–9 years
  \item 10+ years
\end{itemize}

\textbf{A3. Years of professional UX/UI design experience (US-market work)}
\begin{itemize}
  \item < 1 year (\textit{ineligible})
  \item 1–2 years
  \item 3–5 years
  \item 6–9 years
  \item 10+ years
\end{itemize}

\textbf{A4. Have you done professional UX/UI design work for the US market within the past 3 years?}
\begin{itemize}
  \item Yes – based in the US
  \item Yes – based outside the US but working for a US company/client
  \item No (\textit{ineligible})
\end{itemize}

\textbf{\textit{\\***Section B – Professional Background}}

\textbf{B1. Current job title:} \underline{\hspace{2cm}}

\textbf{B2. Company size (most recent)}
\begin{itemize}
  \item 1–10
  \item 11–50
  \item 51–200
  \item 201–1k
  \item 1k–5k
  \item 5k+
\end{itemize}

\textbf{B3. Team setup (multi-select)}
\begin{itemize}
  \item Solo/independent
  \item Small design team (2–5)
  \item Cross-functional squad
  \item Other: \underline{\hspace{2cm}}
\end{itemize}

\textbf{B4. Primary sectors you’ve designed for (multi-select)}
\begin{itemize}
  \item Consumer social/media
  \item E-commerce/marketplaces
  \item Health/wellness
  \item Fintech
  \item Productivity/collab
  \item Media/entertainment
  \item Civic/public sector
  \item Other: \underline{\hspace{2cm}}
\end{itemize}

\textbf{B5. How often do you participate in design reviews/critique/evaluation in your work? (multi-select)}
\begin{itemize}
  \item Rarely
  \item Monthly
  \item Weekly
  \item Multiple times per week
\end{itemize}

\textbf{B6. Typical stage of design review? (multi-select)}
\begin{itemize}
  \item During exploration/research stage
  \item During user flow/wireframe lo-fi stage
  \item During high-fidelity/spec \& delivery
  \item During interactive prototype \& user test stage
  \item All of the above
\end{itemize}

\textbf{B7. Please list any design review methods you are familiar with or have used} (e.g., heuristic evaluation, design critique sessions, Nielsen’s principles, cognitive walkthrough, etc.):\underline{\hspace{2cm}}

\textbf{\textit{\\***Section C – Logistics \& Compensation}}

\textbf{C1. Device for session you plan to use}
\begin{itemize}
  \item Desktop/laptop
  \item Tablet (\textit{Mobile phone not recommended})
\end{itemize}

\textbf{C2. Browser you plan to use}
\begin{itemize}
  \item Chrome
  \item Safari
  \item Firefox
  \item Edge
  \item Other
\end{itemize}

\textbf{C3. Contact email for follow up and compensation:} \underline{\hspace{2cm}}

\subsection{Inter-task Survey}
\label{appendix:inter_task_survey}

This survey contains 3 sections with a total of 19 questions, designed to understand users' empathy, motivation, and perceptions of usability and usefulness when using the \textbf{Self-Proposed} or \textbf{PrivacyMotiv} method to conduct the design review task.\\
Scoring: 7-point Likert (for all questions)\\
\noindent (1: \textit{strongly disagree}; 2: \textit{disagree}; 3: \textit{somewhat disagree}; 4: \textit{neutral}; 5: \textit{somewhat agree}; 6: \textit{agree}; 7: \textit{strongly agree})

\textbf{\textit{\\***Empathy Scale}}

\hspace{0.5em}\textbf{Cognitive Measurement (Perspective-taking, where designers imagine users' thoughts and feelings)}
\begin{itemize}
    \item While reviewing with [METHOD], I actively tried to imagine the feature from vulnerable users’ perspectives.
    \item While reviewing with [METHOD], I considered how specific my UI/UX design might affect users unlike the “typical” target user.
\end{itemize}

\hspace{0.5em}\textbf{Affective Measurement (Empathic Concern, as designers display sincere care for users)}
\begin{itemize}
    \item While reviewing with [METHOD], I felt concerned for users who could be harmed by unintended privacy outcomes.
    \item While reviewing with [METHOD], privacy risks in the design made me feel uneasy on behalf of affected users.
\end{itemize}

\hspace{0.5em}\textbf{Accountability/Action}
\begin{itemize}
    \item While reviewing with [METHOD], I felt responsible to advocate for design changes that reduce privacy risks.
    \item While reviewing with [METHOD], I was motivated to translate concerns into concrete redesign ideas.
\end{itemize}

\textbf{\textit{\\***Motivation Scale}}

\hspace{0.5em}\textbf{Interest/Enjoyment}
\begin{itemize}
    \item While reviewing with [METHOD], I enjoyed identifying privacy issues in the design.
    \item While reviewing with [METHOD], this privacy review felt interesting and engaging.
\end{itemize}

\hspace{0.5em}\textbf{Value/Usefulness}
\begin{itemize}
    \item While reviewing with [METHOD], the activity felt valuable for improving the product’s quality.
    \item While reviewing with [METHOD], I felt the review was useful to my professional goals.
\end{itemize}

\hspace{0.5em}\textbf{Perceived Competence}
\begin{itemize}
    \item While reviewing with [METHOD], I felt capable of identifying potential privacy issues.
    \item While reviewing with [METHOD], I was confident in my ability to propose alternative privacy-aware design ideas.
\end{itemize}

\hspace{0.5em}\textbf{Effort/Importance}
\begin{itemize}
    \item While reviewing with [METHOD], I put a lot of effort into the privacy review.
    \item While reviewing with [METHOD], it was important to me to do well on this review.
\end{itemize}

\textbf{\textit{\\***Method Usability and Usefulness Scale}}

\begin{itemize}
    \item While reviewing with [METHOD], I found the process easy to carry out.
    \item While reviewing with [METHOD], I felt the approach helped me identify privacy issues I might otherwise overlook.
    \item While reviewing with [METHOD], I felt the approach supported me in thinking about vulnerable users.
    \item While reviewing with [METHOD], I felt confident that I was conducting the review effectively.
    \item I would like to conduct this [METHOD] for privacy review in my real life design work.
\end{itemize}

\vspace{2em}
\section{List of Persona Types Used in the Study}\label{appendix:persona_types}
\begin{table}[hbtp]
\centering
\small
\begin{minipage}{\columnwidth}
\begin{itemize}[leftmargin=*, nosep]
  \item Discrimination due to refugee status
  \item Elderly individual with cognitive decline
  \item Individual with chronic illness
  \item Unemployed single parent
  \item Rural subsistence farmer
  \item Survivor of domestic violence
  \item Formerly incarcerated individual
  \item Socially isolated senior
  \item Individual with history of addiction
  \item Disenfranchised ethnic minority
  \item High school dropout
  \item Teen facing online bullying
  \item Child navigating social media
  \item Person with physical disability
  \item Gender non-conforming individual
  \item Low literacy adult
  \item Urban poor teenager
  \item Housing-insecure youth
  \item Person with severe anxiety
  \item Migrant laborer
\end{itemize}
\end{minipage}
\label{tab:persona-types}
\end{table}

\onecolumn

\clearpage
\begin{table*}[t]
\section{Taxonomy of Vulnerability Dimensions (\autoref{tab:vulnerability-taxonomy})}\label{appendix:vulnerability-taxonomy}
\centering
\small
\begin{tabular}{|p{0.28\textwidth}|p{0.64\textwidth}|}
\hline
\textbf{Dimension} & \textbf{Indicators and Examples} \\
\hline
\textbf{1. Structural and institutional exclusion} &
\begin{itemize}
  \item Facing discrimination or exclusion by policies or social structures.
  \item Limited or unequal access to legal protections and public services.
\end{itemize} \\
\hline
\textbf{2. Intersectional identities and characteristics} &
\begin{itemize}
  \item Age-related vulnerability (e.g., children, elderly).
  \item Gender and sexual minorities.
  \item Chronic health conditions or disabilities.
  \item Marginalized due to migrant, refugee, or minority status.
  \item Geographical isolation or environmental risks.
\end{itemize} \\
\hline
\textbf{3. Socioeconomic disadvantage} &
\begin{itemize}
  \item Poverty or unstable financial situations.
  \item Lower levels of education or limited access to learning opportunities.
  \item Inadequate or difficult access to essential resources (e.g., healthcare, housing, nutrition).
\end{itemize} \\
\hline
\textbf{4. Internal and experiential factors} &
\begin{itemize}
  \item Experiencing social stigmatization or marginalization.
  \item Feelings of dependency, isolation, helplessness, or limited autonomy.
\end{itemize} \\
\hline
\textbf{5. Reduced capacity in emergencies and disasters} &
\begin{itemize}
  \item Difficulty in anticipating or preparing for adverse or unexpected events.
  \item Limited ability to effectively cope, resist, or respond during crises.
  \item Prolonged or impaired recovery after an emergency or disaster.
\end{itemize} \\
\hline
\end{tabular}
\caption{Taxonomy of Vulnerability Dimensions Used in Persona Construction}
\label{tab:vulnerability-taxonomy}
\end{table*}

\FloatBarrier
\section{Codebook for Specificity Levels and Thematic Analysis (Table~\ref{tab:spec-levels} \& \ref{tab:theme-defs})}
\begin{table*}[t]
\centering
\small
\setlength{\tabcolsep}{5pt}
\renewcommand{\arraystretch}{1.1}
\begin{tabular}{@{}p{0.11\textwidth} p{0.25\textwidth} p{0.39\textwidth} p{0.21\textwidth}@{}}
\toprule
\textbf{Level} & \textbf{Definition (what it captures)} & \textbf{Coding cues \& examples (abridged)} & \textbf{Typical miscodes (exclude)} \\
\midrule
\textbf{L1 Strategy} & High-level intent, values, or conceptual concerns without committing to specific functions, flows, or UI. & Cues: “The concept is risky,” “Users may regret this,” “The feature encourages surveillance.” Example (P15): “The concept of ‘friend activity’ itself encourages surveillance.” & Purely ethical opinions with no product implications; meta-commentary on the research process. \\

\textbf{L2 Scope} & Concrete \emph{functions or options} are named but not where or when they occur in the UI. & Cues: “Add a privacy setting/option,” “Make it opt-in,” “Allow anonymous posting.” Example (P1): “Add a setting to blur or generalize the user's location.” & Mentions of a specific “button” or “page” (L4–L5); sequencing words like “before/after” (L3). \\

\textbf{L3 Structure} & \emph{Flows, sequence, or timing}: when an action occurs (e.g., before/after a step); persistence or expiration of a state. & Cues: “Before syncing, ask for consent,” “Expires after 6 hours,” “The order is illogical.” Example (P10): “The order of the flow is illogical; users log in to Facebook before setting preferences.” & A bare mention of “add a setting” (L2); naming a specific page/component without a flow context (L4–L5). \\

\textbf{L4 Skeleton} & \emph{Pages, layout, or placement}: where in the UI a control, component, or notice is located. & Cues: “On the Settings page,” “At the top of the feed,” “Group these options together.” Example (P8): “Group all privacy-related options under a single ‘Privacy Settings’ section.” & Microcopy or component styling without placement (L5); pure flow or timing descriptions (L3). \\

\textbf{L5 Visual} & \emph{Specific components or microcopy}: labels, icons, tooltips, banners, colors, or specific wording. & Cues: “The toggle label is unclear,” “Add a tooltip,” “Change the button text,” “Use a warning icon.” Example (P5): “The name ‘Private Session’ is unclear.” & Statements about page layout (L4); statements about flow or timing (L3). \\
\midrule
\multicolumn{4}{p{\textwidth}}{\textit{Decision rule:} Assign the \emph{highest} (most concrete) level explicitly supported by the utterance. For example, “Place a red color highlighted toggle on the Share screen” is coded as L5, not L4.} \\
\bottomrule
\end{tabular}
\caption{Codebook for Specificity Levels (adapted from Garrett’s planes~\cite{garrett2022elements}).}
\label{tab:spec-levels}
\end{table*}


\begin{table*}[t]
\centering
\small
\setlength{\tabcolsep}{5pt}
\renewcommand{\arraystretch}{1.1}
\begin{tabular}{@{}p{0.22\textwidth} p{0.47\textwidth} p{0.27\textwidth}@{}}
\toprule
\textbf{PbD Principle (Theme)} & \textbf{Operational Definition (What it captures)} & \textbf{Examples / Coding Cues} \\
\midrule
\textbf{Proactive not Reactive} & Issues or suggestions related to preventing privacy harms before they occur, often through education, warnings, or systemic safeguards. & Include: “Implement age verification,” “Add a pre-stream checklist to warn users,” “Educate users on consent.” Exclude: Fixes for existing leaks (then \textit{End-to-End Security}). \\

\textbf{Privacy as the Default} & Issues or suggestions concerning the default state of privacy settings, focusing on opt-in versus opt-out models. & Include: “Default sharing is a risk,” “Make sharing opt-in,” “Privacy should be the default.” Exclude: General settings without a default-state focus (then \textit{Respect for User Privacy}). \\

\textbf{Privacy Embedded into Design} & Issues or suggestions related to integrating privacy into the core architecture and functionality, rather than treating it as an add-on. & Include: “Group privacy settings together,” “Reframing a feature's purpose to be less about surveillance,” “The feature's placement is confusing.” \\

\textbf{Full Functionality} & Issues where a privacy control negatively impacts usability, or where the design fails to balance privacy with user experience. & Include: “The warning pop-up is annoying,” “The UI violates mental models,” “The process is too complex.” Exclude: Issues where functionality is not the primary concern. \\

\textbf{End-to-End Security} & Issues related to the full lifecycle of data, including its retention, deletion, and protection from long-term risks like stalking or harassment. & Include: “Historical videos create a permanent record,” “Risk of stalking,” “The app should prohibit screenshots,” “Unclear deletion process.” \\

\textbf{Visibility \& Transparency} & Issues concerning how clearly the system communicates its data practices and the consequences of user actions. & Include: “Unclear language/labels,” “Lack of consent pop-ups,” “The purpose of data collection is not explained,” “No feedback on who can see my profile.” \\

\textbf{Respect for User Privacy} & Issues related to user control, agency, and consent. Focuses on providing users with meaningful choices over their personal information. & Include: “Lack of granular controls,” “All-or-nothing sync,” “Inability to select specific friends,” “No option for anonymity.” \\
\midrule
\multicolumn{3}{p{\textwidth}}{\textit{Assignment rule:} Choose the single principle that best captures the primary privacy issue. If an issue involves both lack of control and unclear text, prefer \textit{Respect for User Privacy} if the core problem is a missing feature, and \textit{Visibility and Transparency} if the core problem is a confusing UI element or text.} \\
\bottomrule
\end{tabular}
\caption{Codebook for thematic analysis (adapted from Privacy by Design principles~\cite{cavoukian2009privacy}).}
\label{tab:theme-defs}
\end{table*}

\clearpage
\setcounter{page}{28}
\begin{figure*}[t]
\section{Prompt pipeline}\label{appendix:prompt}
\subsection{Persona Generation (Figure~\ref{fig:I1_A} \&~\ref{fig:I1_B})}\label{appendix:Input_1}
    \centering
    \includegraphics[width=\textwidth]{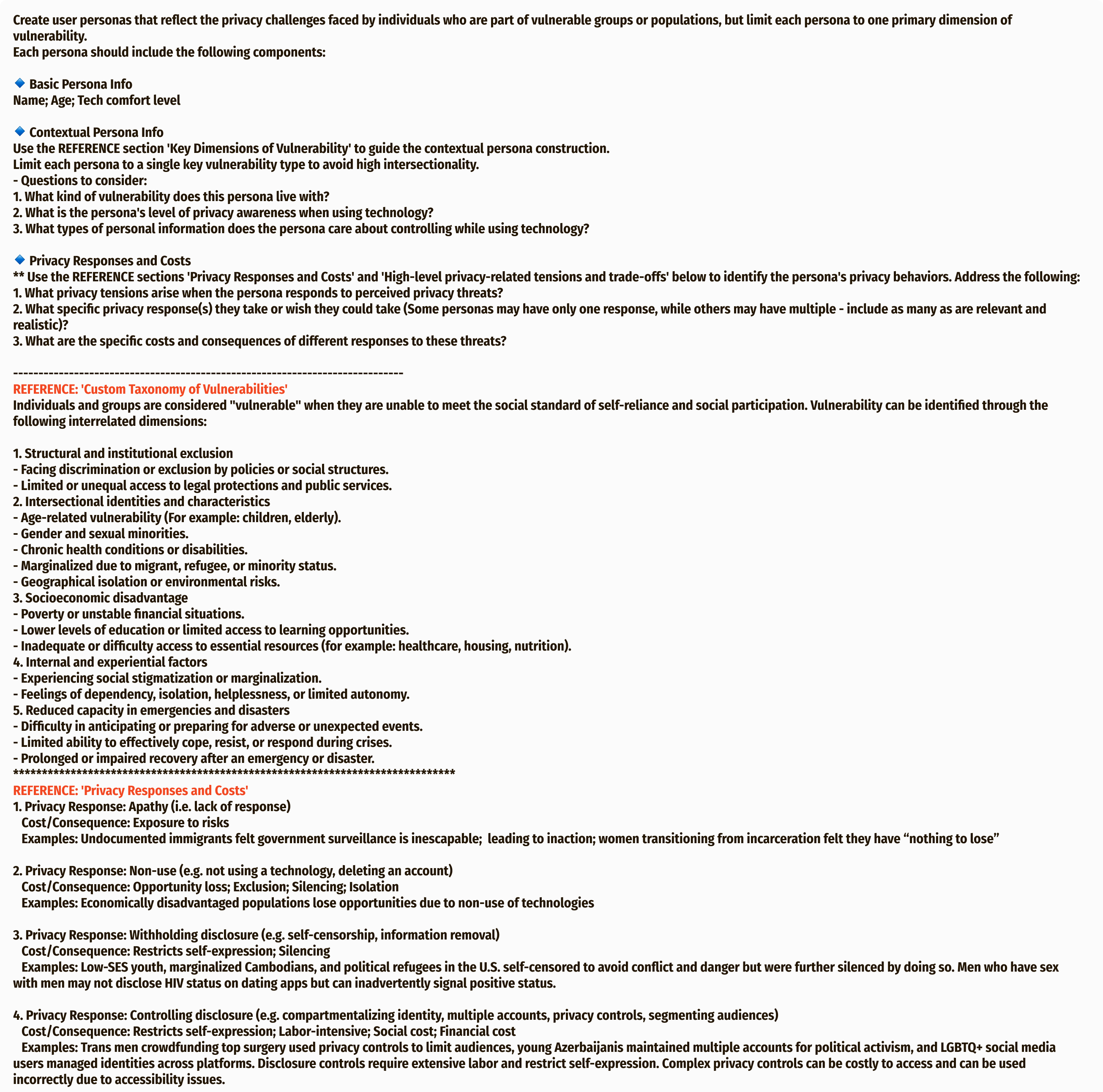}
    \caption{Persona generation prompt}
    \Description{The persona generation prompt used in PrivacyMotiv, showing the system instructions and input structure provided to the LLM}
    \label{fig:I1_A} 
\end{figure*}

\begin{figure*}[hbtp]
    \centering 
    \includegraphics[width=\textwidth]{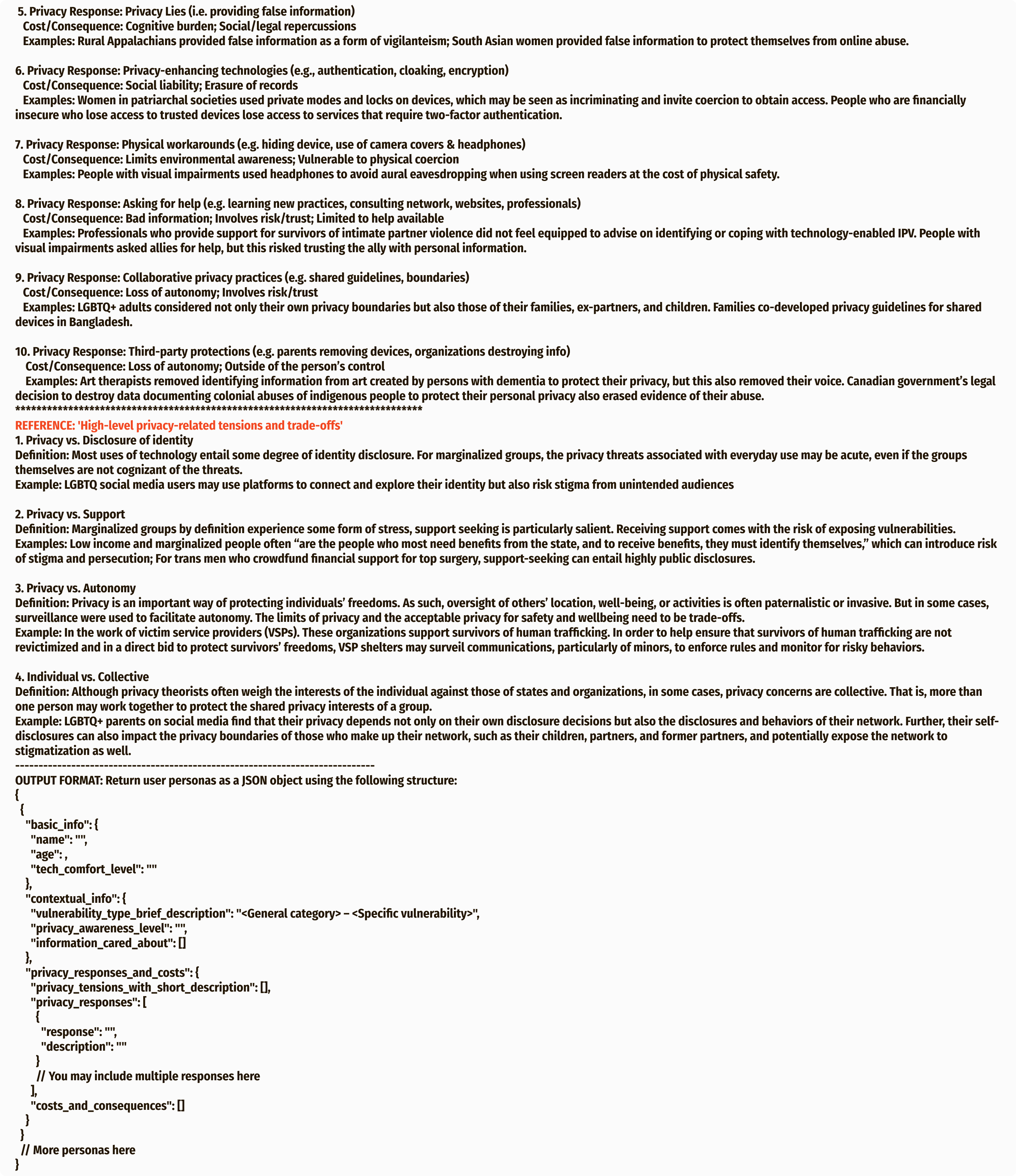}
    \caption{Persona Generation Prompt-Continue} 
    \Description{Continued persona generation prompt, specifying the output format for structured personas}
    \label{fig:I1_B} 
\end{figure*}


\clearpage
\begin{figure*}[t]
\subsection{Privacy-invasive User Journey Narrative (Figure~\ref{fig:I2_M_A}, \ref{fig:I2_M_B}, \ref{fig:I2_N_A}, \ref{fig:I2_N_B})}\label{appendix:Input_2}
    \centering
    \includegraphics[width=\textwidth]{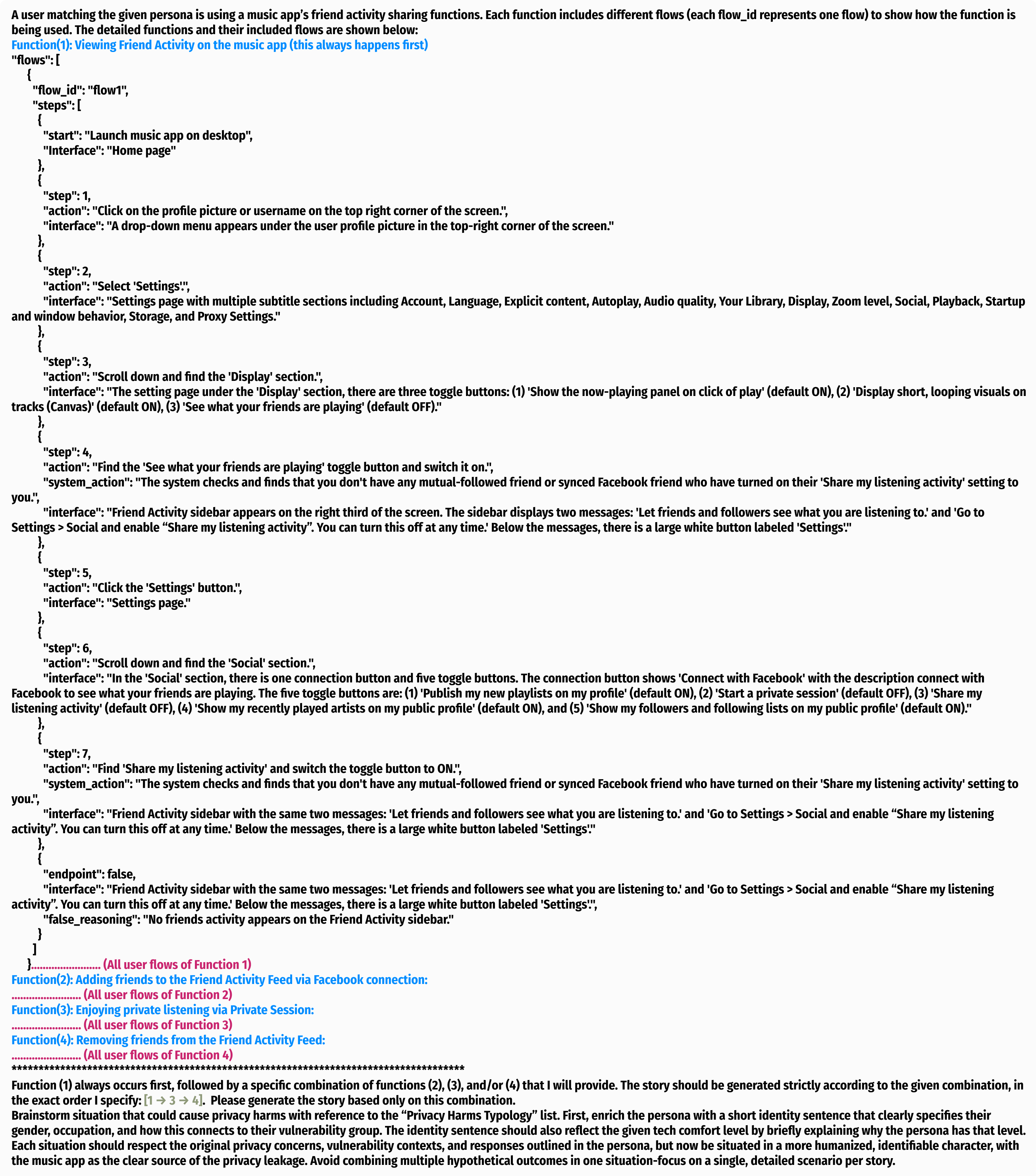}
    \caption{Privacy-Invasive User Journey Narrative Generation Prompt for the WeMusic App}
    \Description{The privacy-invasive user journey narrative generation prompt for the WeMusic app}
    \label{fig:I2_M_A}
\end{figure*}

\begin{figure*}[hbtp]
    \centering
    \includegraphics[width=\textwidth]{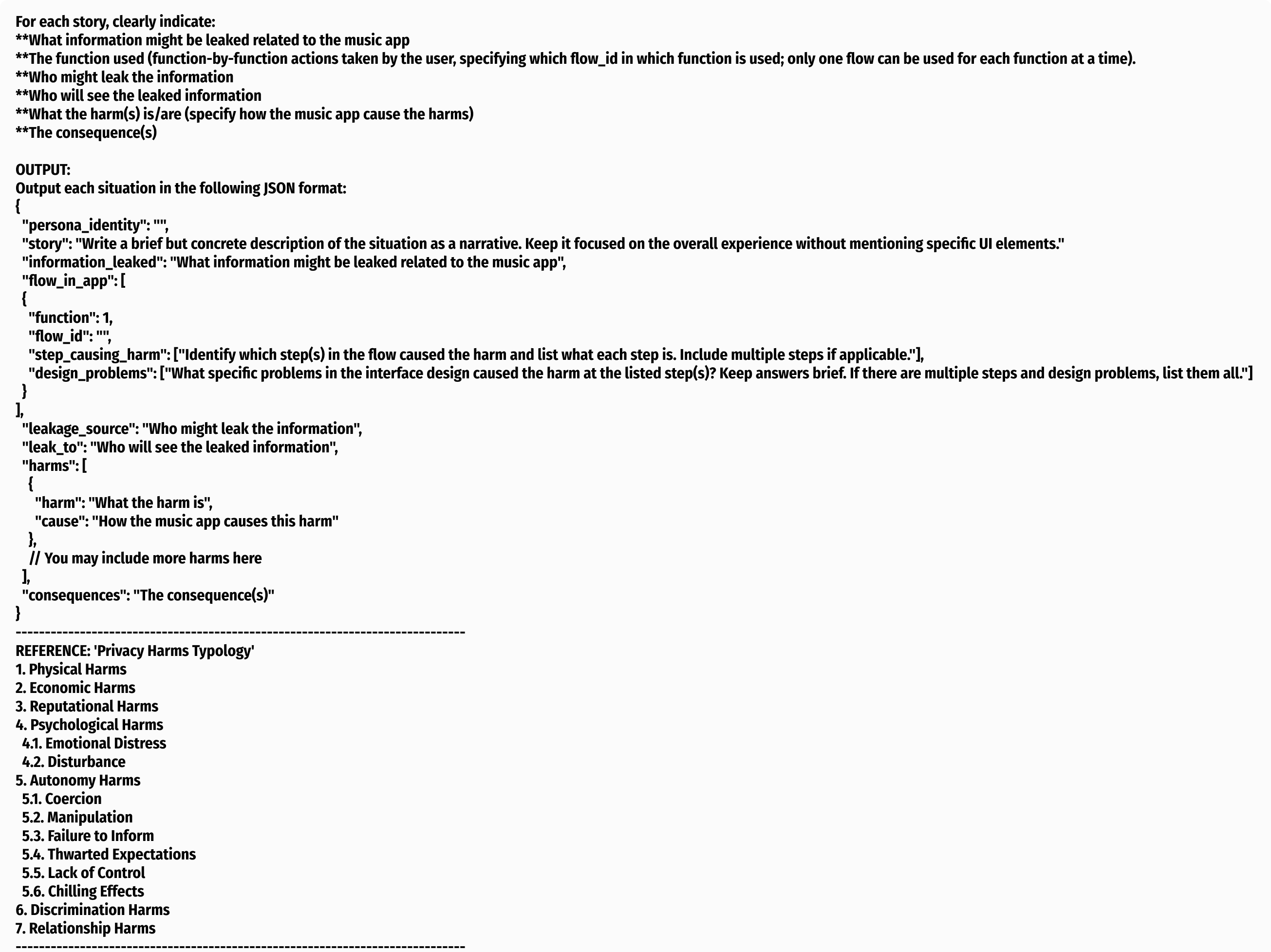}
    \caption{Privacy-Invasive User Journey Narrative Generation Prompt for the WeMusic App-Continue}
    \Description{Continued privacy-invasive user journey narrative generation prompt for the WeMusic app}
    \label{fig:I2_M_B}
\end{figure*}

\begin{figure*}[hbtp]
    \centering
    \includegraphics[width=\textwidth]{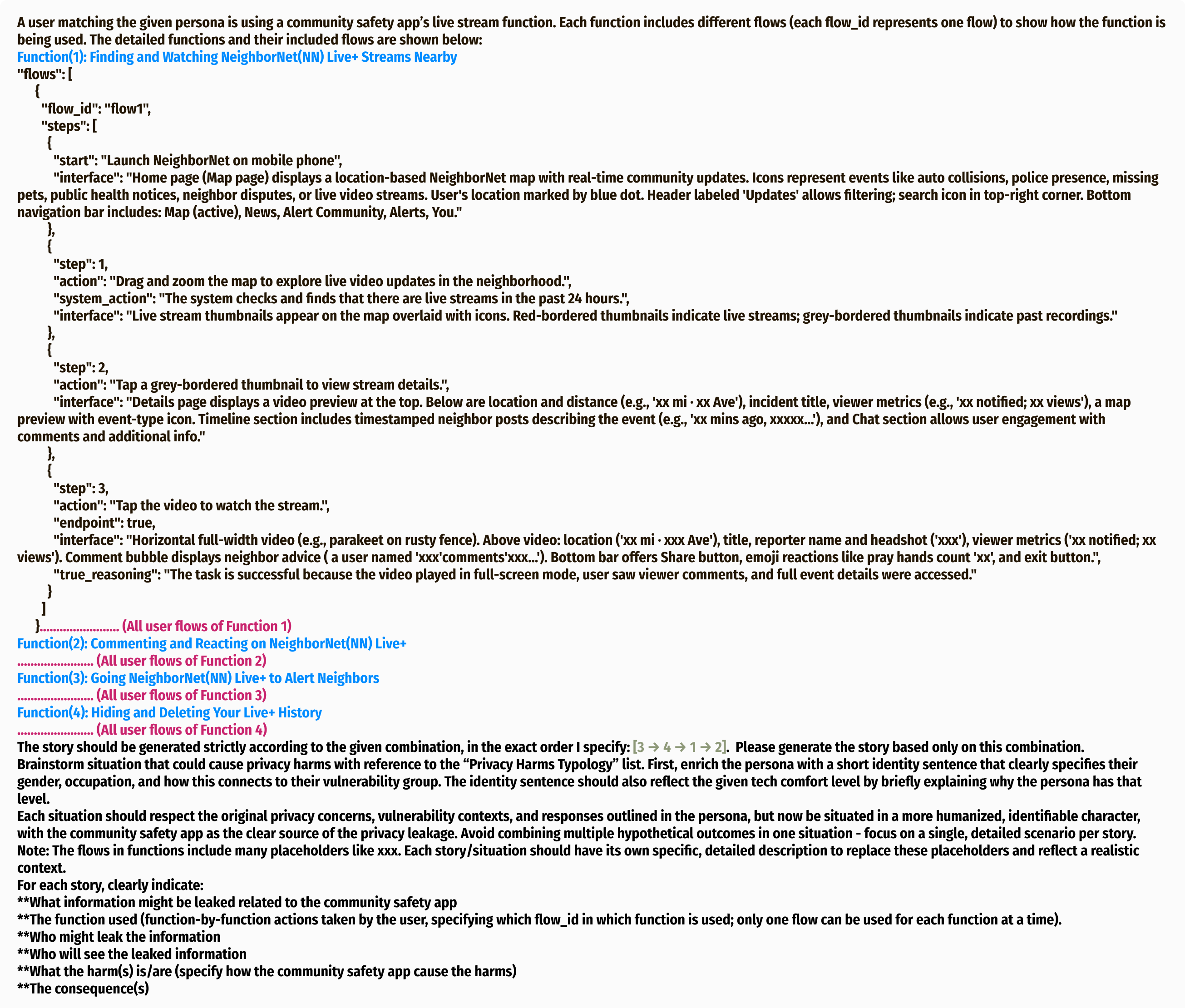}
    \caption{Privacy-Invasive User Journey Narrative Generation Prompt for the NeighborNet App}
    \Description{The privacy-invasive user journey narrative generation prompt for the NeighborNet app}
    \label{fig:I2_N_A}
\end{figure*}

\begin{figure*}[hbtp]
    \centering
    \includegraphics[width=\textwidth]{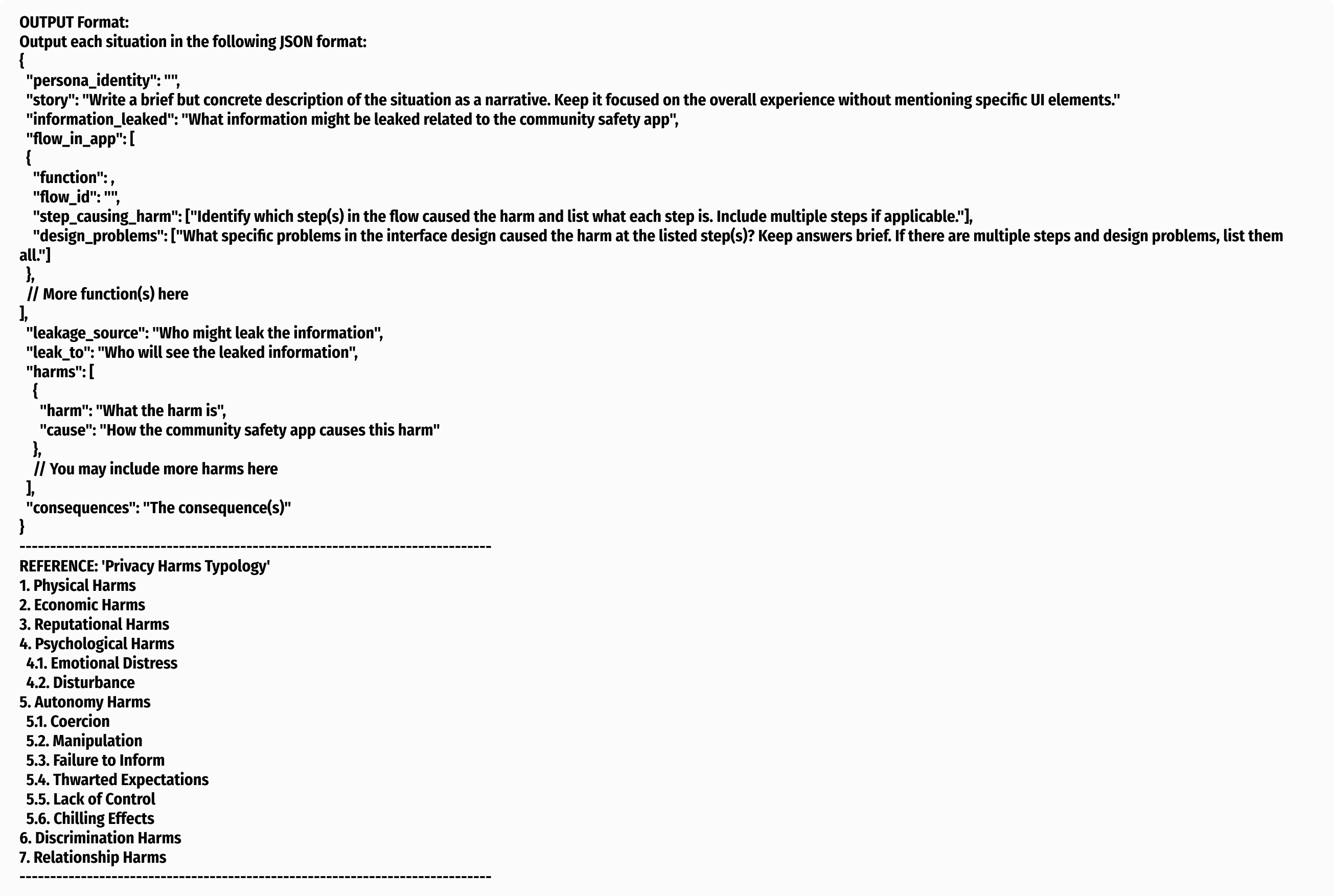}
    \caption{Privacy-Invasive User Journey Narrative Generation Prompt for the NeighborNet App-Continue}
    \Description{Continued privacy-invasive user journey narrative generation prompt for the NeighborNet app}
    \label{fig:I2_N_B}
\end{figure*}

\end{document}